# The degenerate band edge laser: a new paradigm for coherent light-matter interaction

Mehdi Veysi[1], Mohamed A. K. Othman[1], Alexander Figotin[2], and Filippo Capolino[1]

[1]Department of Electrical Engineering and Computer Science, University of California, Irvine, CA 92697, USA
[2]Department of Mathematics, University of California, Irvine, CA 92697, USA

We propose a novel class of lasers based on a fourth order exceptional point of degeneracy (EPD) referred to as the degenerate band edge (DBE). EPDs have been found in Parity-Time-symmetric photonic structures that require loss and/or gain, here we show that the DBE is a different kind of EPD since it occurs in periodic structures that are lossless and gainless. Because of this property, a small level of gain is sufficient to induce single-frequency lasing based on a synchronous operation of four degenerate Floquet-Bloch eigenwaves. This lasing scheme constitutes a new paradigm in the light-matter interaction mechanism that leads also to the unprecedented scaling law of the laser threshold with the inverse of the fifth power of the laser-cavity length. The DBE laser has the lowest lasing threshold in comparison to a regular band edge laser and to a conventional laser in cavities with the same loaded quality ($Q$) factor and length. In particular, even without mirror reflectors the DBE laser exhibits a lasing threshold which is an order of magnitude lower than that of a uniform cavity laser of the same length and with very high mirror reflectivity. Importantly, this novel DBE lasing regime enforces mode selectivity and coherent single-frequency operation even for pumping rates well-beyond the lasing threshold, in contrast to the multifrequency nature of conventional uniform cavity lasers.

## I. INTRODUCTION

Demonstration of a low-threshold laser operating at a single frequency is an important quest in the optical and physical sciences. In this regard, the use of periodic structures with engineered dispersion diagram is a popular and effective way to enhance the interaction between the gain medium and the electromagnetic wave and therefore tailoring the lasing characteristics of active structures. In the last decades, photonic-crystals-based optical devices and distributed feedback (DFB) lasers have demonstrated inimitable features and high performance due to their unprecedented dispersion characteristics, high quality ($Q$) factors, and field enhancement properties [1–6]. In particular, increasing the $Q$-factor of photonic-crystal-based cavities results in a significant reduction of the lasing threshold [2–4,7–11]. Therefore, many techniques have been proposed to enhance the $Q$-factor of photonic-crystal-based cavities for engineering light sources, such as introducing a small disorder or a defect into the crystal [7,12], using photonic-crystal heterostructures [9,13], and by locally modulating the width of the photonic-crystal waveguide [14,15]. Recent advances in developing optical lasers rely on engineering the response of the cavity structures by employing plasmonic nanocavities [16–18], photonic band-edges [19–25], Parity-Time (PT)-symmetry breaking [26–32], or by exploiting unique structural topologies including metamaterials [33–36] and metasurfaces [37,38].

In this paper, we propose a novel class of single-frequency lasers made of a cavity with degeneracy of four Floquet-Bloch eigenwaves coherently interacting with an active medium. Such a degeneracy is found in periodic structures whose dispersion relations develop points of degeneracy at which state eigenvectors representing Floquet-Bloch eigenwaves coalesce [21,22,39]. Those points in the spectrum of the "cold" periodic structure ("cold" refers to the absence of the gain media) are associated with a regular band edge (RBE), a stationary inflection point (SIP) [40,41], or a degenerate band edge (DBE) [22,23], resulting in a second, third, or fourth order degeneracy of Floquet-Bloch eigenwaves (in both eigenvalues and eigenvectors), respectively. We refer to such points as *exceptional points of degeneracy (EPDs)*.

EPDs have been commonly associated with the presence of gain and/or loss and often related to Parity-Time (PT)-symmetry [26–32]. However here we point out that the EPD may be induced in electrodynamical systems also in absence of gain and losses. In gain and loss balanced systems (like in systems with PT-Symmetry), EPDs occur in the parameter space of the system described by the evolution of their eigenmodes either in time (for coupled resonators such as those in such as those in [31,32,41]) or in space (for coupled waveguides such as those in [42,28,43,44]). On the other hand, EPDs also realizable in spatially-periodic media supporting Floquet-Bloch waves such as photonic crystals and periodic waveguides exhibiting RBEs, SIPs, and DBEs, *in absence of gain and loss*. It is then important to emphasize that indeed both cases of (i) gain and loss induced EPDs, and (ii) periodicity induced EPDs, follow the same mathematical fundamental theory of degenerate operators (see page 63-67 in [45]), i.e., the eigenvalues and eigenvectors characteristics in both cases lead to a Jordan Block degeneracy as described in [22,43–45]. In this paper we investigate the space-evolution of guided eigenwave in periodic waveguides (i.e., along the *z*-direction) at and near EPDs, that would result in unique lasing features.

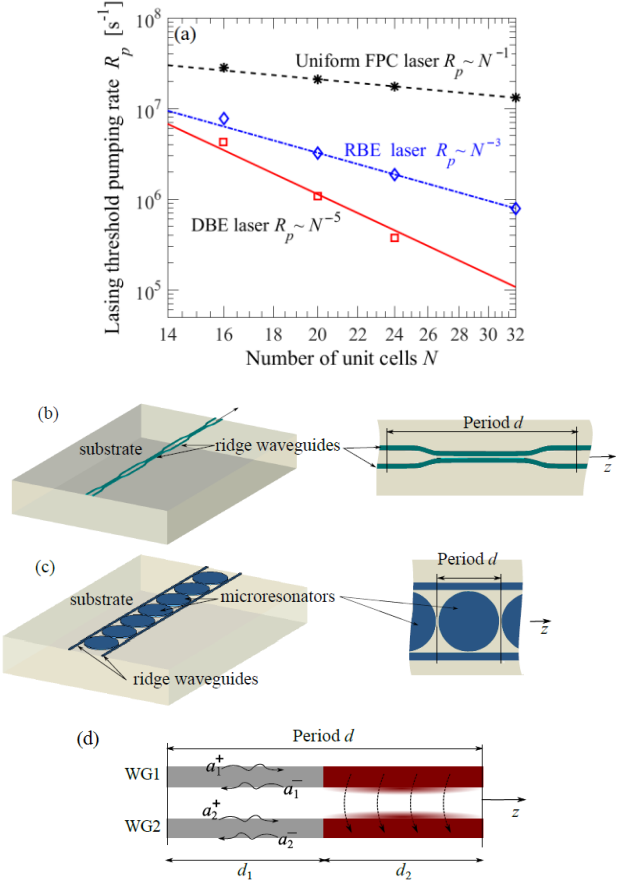

FIG. 1. (a) Scaling of the lasing threshold pumping rate of three regimes of laser operation based on three types of cavities varying as a function of the cavities' lengths (number of unit cells $N$). The EPD-based laser proposed in this paper demonstrates the lowest lasing threshold as well as a unique scaling of its threshold versus the active medium length as $N^{-5}$. The trend is for lossless cavities, i.e., no dissipative losses in the waveguide (methods, analysis, and impact of loss are elaborated in Sec. II through Sec. V). (b-d) Geometries of two basic "cold" (cold refers to the absence of the gain medium) optical ridge waveguides that exhibit fourth order degeneracy (i.e., a DBE). The ridge waveguides here are periodically coupled using different coupling mechanism: (b) proximity coupling, and (c) using optical microresonators (e.g., microrings or microdiscs). Such waveguide geometries with different coupling mechanisms are represented here by an equivalent waveguide model with structured periodic unit cell. (d) A unit cell of the periodic waveguide specifically considered in this paper to prove the proposed concepts. It has a length $d$ and is composed of one uncoupled section (gray color) and another coupled section (dark red color) with lengths of $d_1$ and $d_2$ respectively, designed to exhibit a DBE at optical communication wavelength of $\lambda_d = 1550\,\mathrm{nm}$.

In particular we focus on the DBE degeneracy [22,39,46,47] which arises when *four* Floquet-Bloch eigenwaves coalesce in spatially-periodic structures supporting multiple polarization states that are periodically mixed. It features a frozen-wave resonance relying on a fundamental property of EPDs that causes eigenwave solutions inside a periodic structure to diverge, leading to giant field enhancement [22,47]. However, it is important to stress that the DBE, which is a fourth order EPD, occurs in a passive and lossless system, i.e., without the need of gain or loss. Some DBE characteristics have been shown to occur at optical frequencies using perturbed coupled silicon waveguides [48,49], or a chain of ring resonators coupled to a waveguide [50]; as well as in metallic waveguides at microwaves [39]. There have been also significant efforts in analysis, design, and experimental realization of the DBE structures and its slow-wave properties at both microwave [51–53] and optical frequencies [54,55]. It is important to point out that even though DBE is a precise EPD condition occurring in lossless waveguides, experimental studies made both at microwaves [53] and optical frequencies [50,51] have confirmed the existence of features associated with the DBE in the presence of losses and fabrication tolerances. Indeed, the robustness of DBE features against perturbations due to possible fabrication tolerances that may arise during fabrication was demonstrated for DBE CROWs in [50]. The DBE has led to observing giant gains in optical cavities [47]; however here we leverage a general EPD concept, for the first time, to propose the new regime of lasing, resulting in low-threshold and single frequency operation of the degenerate band edge laser.

In previous work [47], an analysis of an ideal multilayer anisotropic medium with ideal gain (not represented with rate equations in a multilevel energy setup as done here) is carried out using the transfer matrix method which has led to a new route for possible gain enhancement in cavities with DBE. The analysis in [47] has resulted in an oscillation threshold of such DBE active cavities that scales as $N^{-5}$, where $N$ is the number of unit cells as seen from Fig. 1. In this paper we carry the first comprehensive analysis of the proposed DBE laser : (i) we demonstrate a special feature of lasing-mode selection in the DBE laser that leads to a single frequency operation[the analysis of the lasing threshold is elaborated in Sec. II through Sec. IV]; (ii) The study included time domain simulations of electromagnetic fields and evolution of rate equations describing gain arising from a multilevel energy system; (iii) we further show that the proposed DBE laser features a significantly lower lasing threshold compared to a conventional RBE laser [24] and a uniform Fabry-Perot cavity (FPC) while having the same gain medium, length, and same total loaded $Q$-factor [see Fig. 1(a)]. (iv) The low lasing threshold of the proposed DBE laser, which can be up to two orders of magnitude lower threshold than conventional lasing cavities as shown in Fig. 1(a), is ascribed to the enhanced interaction between the gain media and the cavity featured by all the four degenerate Floquet-Bloch eigenwaves at the DBE wavelength as will be explained throughout the paper. (v) We also demonstrate that the DBE laser operates without the need of cavities mirrors and the threshold is independent of mirror reflectivity as shown in Sec. IV. These findings are especially valuable for further developing optical devices based on degeneracy properties. Note that although the DBE is a slow-light phenomenon occurring strictly in a lossless periodic waveguide, we propose the DBE lasing regime in a fully-coupled system composed of a "cold" DBE cavity and a non-linear gain medium (modeled by (4) in Sec. II.B) for which the interaction is investigated using time-dependent non-linear evolution equations.

The concepts discussed here set forth a new paradigm shift in lasers sources since they are based on a novel interaction regime between light and active material. Typically, laser sources operate based on conventional Fabry-Perot cavity resonances and that causes challenges especially in the semiconductor laser realm. However, here we employ a fourth-order eigenwave degeneracy (i.e., they form a *single* degenerate eigenmode) to enforce lasing mode-selectivity and conceive a class of low-threshold lasers whose threshold is independent of mirror reflectivity.

The layout of the paper is organized as follows. First, we show possible implementations of "cold" coupled optical waveguides whose Floquet-Bloch eigenwaves support the DBE in Sec. II.A. We then theoretically investigate the DBE laser based on evolution equations for waves in coupled waveguides that account for spatial periodicity as well as loss and gain in Sec. II.B and Appendix A. The properties of the "cold" DBE structure as well as the steady state gain medium response are introduced and investigated in Sec. II.C and II.D, respectively. We then report the evolution of the lasing action inside the DBE laser using finite-difference time-domain (FDTD) algorithm as well as the lasing threshold analysis [in Fig. 1(a)] in Sec. III and Appendix B. Finally, we demonstrate the effect of losses on the loaded $Q$-factor and provide comparisons between conventional lasers and the proposed DBE laser in terms of lasing threshold in sec. IV.

## II. LASER THEORY IN COUPLED WAVEGUIDES WITH EPDS

In this paper, we propose and investigate a new class of EPD lasers. Our proposed laser operates near the DBE which guarantees coherent single frequency oscillation as well as low threshold.

### A. Pair of Periodic Coupled Waveguides with Four Degenerate Eigenwaves

Among the many possible lossless optical coupled waveguide geometries that may exhibit DBE, i.e., an EPD caused by coalescence of four Floquet-Bloch eigenwaves into one degenerate eigenwave, we illustrate in Fig. 1(b)-(c) two representative waveguide examples. The dispersion relation with DBE is shown in Fig. 2. The optical waveguide in Fig. 1(b) is composed of periodic segments of coupled and uncoupled optical waveguides, with period $d$. Alternative coupling mechanism could be also realized through optical resonators as couplers to the waveguide [as in Fig. 1(c)]. The reason for utilizing optical resonators (such as microrings or microdiscs [56–58]) is their ubiquitous use in lasers due to their high loaded $Q$-factor and ease of fabrication. In the example shown in Fig. 1(c) the DBE cavity is made of waveguides coupled to a chain of coupled resonator optical waveguide (CROW) as shown in [50]. (The conventional CROW topology would support only an RBE [59–61].) Furthermore, the DBE CROW proposed in [50] was shown to exhibit remarkably high $Q$-factor resonances near the DBE that are robust against disorders and perturbations. Note that other geometries can also be designed and implemented using various semiconductor photonics technologies [46]. The geometries shown in Fig. 1(b)-(c) are simply represented using their equivalent coupled waveguide model in Fig. 1(d), constituting WG1 and WG2 that is the model used in the rest of the paper. Such periodic waveguides have a dispersion diagram as in Fig. 2, exhibiting the DBE at $f_d = \omega_d/(2\pi) \approx 193.4\,\text{THz}$, i.e., at $\lambda_d = 1550$ nm, using a pair of periodic coupled waveguides as in Fig. 1(d) with parameters in Appendix C. Note that the dispersion relation in the vicinity of DBE frequency is approximated by $(\omega_d - \omega) \approx h(k - k_d)^4$ where $k_d = \pi/d$ is the wavenumber at the DBE., and $h$ is a geometry dependent parameter. At any $\omega$ the Floquet-Bloch eigenwaves in a pair of coupled periodic waveguides comprise four $k$ wavenumbers, associated to four eigenvectors that are, in general, mutually independent. However, at a DBE frequency the four Floquet-Bloch eigenwaves coalesce, in wavenumber and eigenvectors. A consequence of Floquet-Bloch eigenwave degeneracy is that independent basis of wave propagating must come in the form of generalized eigenvectors at the DBE (see Ch. 7.8 in [63] as well as [64,65]). At the DBE, waves associated to those generalized eigenvectors grow linearly, quadratically and cubically with the $z$ coordinate; having vanishing group velocity yet still satisfying Maxwell's equations [64,66,67]. In addition, finite structures made of such periodic coupled waveguide experience unusual FPC resonances; compared to uniform FPCs [47]. The DBE laser proposed here oscillates based on the FPC resonance closest to the DBE frequency and therefore enables a remarkable low-threshold single-frequency lasing as it will be shown in Secs. III and IV.

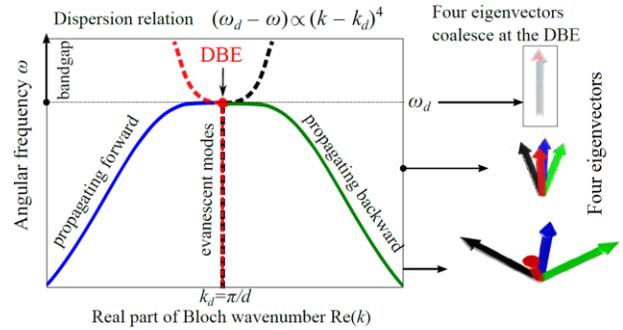

FIG. 2. The dispersion diagram of the "cold" (i.e., in absence of the gain medium) coupled-mode waveguides in Fig. 1 near the DBE, showing the real part of the Bloch wavenumber $k$ versus angular frequency. A distinguished characteristic of the DBE condition is that its dispersion curve follows the asymptotic trend $(\omega_d - \omega) \propto (k - k_d)^4$ near the DBE angular frequency $\omega_d = 2\pi f_d$, $f_d \approx 193.4\,\text{THz}$. The right panel is a representation of eigenvectors evolution with frequency in the periodically-coupled waveguide. Note that there are four eigenvector solutions corresponding to two propagating eigenwaves (forward with Re($k$)<$\pi/d$ and backward with Re($k$)>$\pi/d$) in solid lines and two evanescent eigenwaves in dashed lines. At the DBE frequency, these four eigenvectors coalesce into one degenerate eigenvector as shown in right panel of the figure. Also, there exists a band gap for frequencies higher than that of the DBE.

Note that the concepts exposed in this paper are general and applicable to any waveguide geometry exhibiting a DBE. To illustrate the DBE laser concept however we specifically refer to an illustrative example of periodic waveguides modeled as a cascade of coupled and uncoupled waveguides as in Fig. 3 (details in the Appendices) with coupled-wave equations described as follows.

### B. Time Domain Formulation of The DBE Laser Action

We describe in detail the theory of coupled waves propagating (along the $z$-direction) in coupled waveguides with EPDs. We assume that each of the two waveguides has only a single propagation eigenwave (in each of the $\pm z$-direction), so the coupled waveguide system has two propagating eigenwaves (in each direction). Here coupling between the two waveguides (WG1 and WG2) is mediated phenomenologically through coupled differential wave equations [as in (1)] involving wave amplitudes. (Note that this approach is known in the radio frequency as coupled transmission lines [68] and it is inherently related to the conventional coupled-mode theory [69–74] widely used for optical systems). We again point out that the lasing regime proposed here is based on a fully-coupled system modeled by non-linear evolution equations in time, describing the dynamics of electromagnetic waves in the DBE cavity that incorporates gain material. Discussions about gain enhancement properties for slow-light waveguides can be found elsewhere, for instance in [24,25,47]. We stress that, strictly speaking, the presence of gain or loss detunes the system away from the mathematical DBE condition as we have previously investigated in other cases [75,65,76,47]. However important field properties of the special DBE degeneracy are retained when gain is not high. In addition, our evolution equations correctly take into account gain in a slow-light system and would provide proper results also for large gain.

We consider two traveling waves, in WG1 and WG2, that are described by their spatiotemporal amplitudes $a_1^+(z,t)$ and $a_2^+(z,t)$. The two waves propagate along the coupled waveguide in the $+z$-direction. Their amplitudes are normalized in such a way to represent power waves as pointed out in [77], so that $S^+(z,t) = \left(a_1^+(z,t)\right)^2 + \left(a_2^+(z,t)\right)^2$ is the instantaneous power flux in the coupled waveguides along the $+z$-direction. As such the wave amplitudes are conveniently expressed using a two-dimensional vector $\mathbf{a}^+(z,t) = \begin{bmatrix} a_1^+(z,t) & a_2^+(z,t) \end{bmatrix}^T$ where superscript $T$ stands for transpose. From here onward, bold symbols denote vectors, while bold double-underlined symbols denote 2×2 matrices and bold single-underlined symbols denote 4×4 matrices.

The space-time evolution equations for the wave amplitudes $\mathbf{a}^+(z,t)$ in a uniform, lossless coupled waveguide are given by

$$c\frac{\partial}{\partial z}\begin{pmatrix} a_1^+(z,t) \\ a_2^+(z,t) \end{pmatrix} = -\underline{\underline{\mathbf{n}}}\frac{\partial}{\partial t}\begin{pmatrix} a_1^+(z,t) \\ a_2^+(z,t) \end{pmatrix}, \quad \underline{\underline{\mathbf{n}}} = \begin{bmatrix} n_{11} & n_m \\ n_m & n_{22} \end{bmatrix}, \quad (1)$$

where $c$ is the speed of light in vacuum, $n_{11}$, $n_{22}$ and $n_m$ are effective refractive indices of the coupled waveguide, respectively, and $\underline{\underline{\mathbf{n}}}$ is a 2×2 matrix whose eigenvalues are the effective refractive indexes of the propagating waves in the coupled waveguides. Note that the coupling between the two waveguides is present through the off-diagonal entries $n_m$ of the matrix $\underline{\underline{\mathbf{n}}}$. Indeed, when $n_m = 0$ the two waveguides are uncoupled and the space-time evolution of the waves in the system described by (1) turns into two uncoupled equations whose solutions are the natural propagating eigenwaves in WG1 and WG2. When WG1 and WG2 are coupled, the two eigenwave solutions are found by solving the coupled-wave equation (1). Thanks to reciprocity, the eigen solutions of (1) obey the symmetry $t \rightarrow -t$. Therefore, independent eigenwaves may also propagate in the negative $z$-direction, and their amplitudes are denoted by a two dimensional vector $\mathbf{a}^-(z,t) = \begin{bmatrix} a_1^-(z,t) & a_2^-(z,t) \end{bmatrix}^T$. The evolution of those wave amplitudes is also obtained from (1) through the transformation $a^+ \rightarrow a^-$ and $t \rightarrow -t$. Accordingly, we write the evolution equations for both $\mathbf{a}^+$ and $\mathbf{a}^-$ in a matrix form for a uniform waveguide viz,

$$c\frac{\partial \mathbf{a}^+(z,t)}{\partial z} = -\underline{\underline{\mathbf{n}}}\frac{\partial \mathbf{a}^+(z,t)}{\partial t} \ ; \quad c\frac{\partial \mathbf{a}^-(z,t)}{\partial z} = \underline{\underline{\mathbf{n}}}\frac{\partial \mathbf{a}^-(z,t)}{\partial t}. \quad (2)$$

In (1), that applies to a lossless system, the matrix $\underline{\underline{\mathbf{n}}}$ is purely real and symmetric (so it is Hermitian). The entries of $\underline{\underline{\mathbf{n}}}$ are real valued and are associated to the refractive index of the coupled waveguide system. The matrix $\underline{\underline{\mathbf{n}}}$ appears as a simple multiplier because in a frequency domain description we neglect material and waveguide frequency dispersion. This is a valid approximation since we investigate lasing action in a narrow frequency range given by the emission spectrum of the active atoms. In general, one may construct a first order evolution equation in the waveguide by assuming a state vector $\hat{\mathbf{\Psi}}(z,t)$ comprising the wave amplitudes, that evolves in the lossless coupled waveguide as

$$\frac{\partial \hat{\mathbf{\Psi}}(z,t)}{\partial z} = -\hat{\underline{\mathbf{M}}} \frac{\partial \hat{\mathbf{\Psi}}(z,t)}{\partial t}, \text{ and } \hat{\mathbf{\Psi}}(z,t) = \begin{bmatrix} \mathbf{a}^+(z,t) \\ \mathbf{a}^-(z,t) \end{bmatrix}, \quad (3)$$

where the 4×4 matrix $\hat{\underline{\mathbf{M}}}$ is a block-diagonal system matrix comprised of the matrix $\underline{\underline{\mathbf{n}}}$ for each uniform waveguide segment, as given in Appendix A.

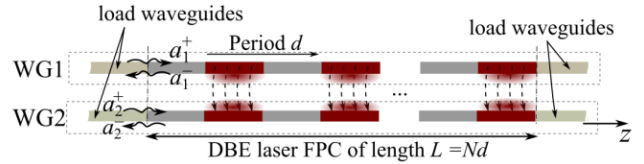

FIG. 3. Schematic representation of an FPC with DBE composed of a finite number of periodic unit cells, each made of a coupled and an uncoupled section, and terminated at the two ends by output ("load") waveguides. The FPC DBE resonance frequency $f_{r,d}$ is very close to the DBE frequency $f_d$.

It is important to point out that the coupling mechanism and the transfer of power between the coupled waveguides are well-understood from the evolution of the coupled waves as described above. The aforementioned analysis resembles the coupled-mode theory for optical waveguides [69–74]. However, we recall that our approach is in fact equivalent to transmission line theory [78] for optical waveguides where we incorporate gain and loss in the temporal evolution equations as we elucidate in the following.

It is convenient to study *total* electromagnetic fields inside the cavity by resorting to electric and magnetic fields' amplitudes, namely by two-dimensional vectors $\mathbf{E}(z,t) = \begin{bmatrix} E_1(z,t) & E_2(z,t) \end{bmatrix}^T$ and $\mathbf{H}(z,t) = \begin{bmatrix} H_1(z,t) & H_2(z,t) \end{bmatrix}^T$, respectively, where the vector components represent amplitudes of the total fields in WG1 and WG2. The two-dimensional vectors $\mathbf{E}$ and $\mathbf{H}$ are related to the wave amplitudes $\mathbf{a}^+$ and $\mathbf{a}^-$ as shown in (A2) in Appendix A, by resorting to the concept of characteristic impedance of the waveguides as discussed in Appendix A (a procedure described in coupled transmission line theory [68]). The reason for adopting this $\mathbf{E}$ and $\mathbf{H}$ field representation is that it is straightforward to include losses and gain in the total field formulation. Such coupled-wave formulation can be then readily characterized using conventional FDTD implemented by a standard Yee [79] algorithm that has been extensively studied for transmission lines [80] as well as for formulations based on both E and H fields [80–82]. As done in Ref. [22,47], for example, the state vector that describes the total field amplitudes is denoted by $\mathbf{\Psi}(z,t) = \begin{bmatrix} \mathbf{E}^T(z,t) & \mathbf{H}^T(z,t) \end{bmatrix}^T$ and it is related to normalized wave amplitude state vector $\hat{\mathbf{\Psi}}(z,t)$ through a matrix transformation (see Appendix A). The space-time evolution of the total field amplitude state vector $\mathbf{\Psi}(z,t)$ in the periodic waveguide is constructed in a similar fashion to (3) and is provided in Appendix A, (A6).

We proceed by generalizing of the above analysis to periodic structures made of sections of uniform coupled waveguides in the presence of gain and loss. The nonlinear gain is provided by an externally-pumped active medium and is incorporated into the analysis through the polarization density amplitudes $P_1(z,t)$ and $P_2(z,t)$. These polarization densities represent the effective polarization field amplitudes induced in WG1 and WG2 because of the gain medium and depend on how much the field distributions associated to the amplitudes $E_1(z,t)$ and $E_2(z,t)$ overlap with the gain medium. Hence, generalizing (A6) to the case of gain and loss one obtains

$$\frac{\partial \mathbf{\Psi}(z,t)}{\partial z} = -\underline{\mathbf{M}}(z)\frac{\partial \mathbf{\Psi}(z,t)}{\partial t} + \begin{pmatrix} \underline{\mathbf{0}} \\ \underline{\underline{\mathbf{s}}} \end{pmatrix}\frac{\partial \mathbf{P}(z,t)}{\partial t} - \begin{pmatrix} \underline{\underline{\mathbf{0}}} & \underline{\underline{\mathbf{0}}} \\ \underline{\underline{\gamma}} & \underline{\underline{\mathbf{0}}} \end{pmatrix}\mathbf{\Psi}(z,t) \quad (4)$$

where $\underline{\mathbf{M}}(z)$ is the system matrix of the lossless and gainless coupled waveguide given in (A7). The two-dimensional polarization density amplitude vector $\mathbf{P}(z,t) = \begin{bmatrix} P_1(z,t) & P_2(z,t) \end{bmatrix}^T$ represent the polarization density in WG1 and WG2 induced by the transition of the active atoms. Therefore $\mathbf{P}(z,t)$ accounts for the active material (gain), and $\underline{\underline{\mathbf{s}}}$ is a 2×2 gain coupling matrix that represents the interaction strength of the gain media and the coupled waveguide fields. The 2×2 matrix $\underline{\underline{\gamma}}$ is a per-unit-length dielectric loss parameter, also given in Appendix A. The atomic transitions occur within a simplified, yet realistic, four-level energy atomic system shown in Fig. 4. Accordingly, the time evolution for the polarization density amplitudes in WG1 and WG2 at the presence of a forcing electric field is described by the homogeneously broadened Lorentzian oscillator model and obtained as [82–85]

$$\left[\frac{\partial^2}{\partial t^2} + \Delta\omega_e \frac{\partial}{\partial t} + \omega_e^2\right]\mathbf{P}(z,t) = -\sigma_e \Delta N \underline{\underline{\mathbf{s}}} \mathbf{E}(z,t), \quad (5)$$

where

$$\sigma_e = 6\pi\varepsilon_0 c^3 / (\tau_{21}\omega_e^2), \quad (6)$$

and $\tau_{21}$ is the photon lifetime of the transition between the second and the first energy states (i.e., between energy levels 2 and 1), $e$ and $m$ are the charge and the mass of an electron, $\varepsilon_0$ is the free-space permittivity, $\omega_e$ is the angular frequency of emission (chosen for our specific implementation at $\lambda_e = 1550 \text{ nm}$), and $\Delta\omega_e$ is the full width at half maximum line-width of the atomic dipolar transition. The 2×2 matrix $\underline{\underline{s}}$ in (5) is given in Appendix C and indicates which of the two waveguides is involved in the coupling between gain media and electric field amplitudes. Here, its components are simply taken as 1 or 0, though it could take into account also other confinement factors. The atomic system has four energy levels with atomic population densities of $N_0$, $N_1$, $N_2$, and $N_3$ and $\Delta N$ in (5) is also the population density difference between the first and the second energy levels (i.e., $\Delta N = N_2 - N_1$) [83,84]. The time- and space-dependent population density $N_j(z,t)$ at each energy level $j = 0,1,2,3$ is obtained from the nonlinear rate equations provided in the Appendix A. Here, we assume that the total active material density $\left(\text{i.e., } N_T = \sum_{j=0}^{3} N_j(z,t)\right)$ is uniformly distributed and is invariant with space and time. The cavity is optically-pumped with pumping rate $R_p$ transferring active atoms from the ground state (0th level) into the highest energy level (3rd level). The lasing action occurs when the atomic transition from the second level to the first one is slow and radiative, leading to a population inversion (i.e., lasing condition is when $\tau_{21} > \tau_{32}, \tau_{10}$ which results in $N_2 > N_1$). Here, we assume that the active material is Erbium ($Er^{3+}$) and it is doped into the substrate. The pumping rate $R_p$ is a tunable parameter and can be varied by the external pump intensity in a real experiment. Therefore, we fully characterize the laser cavity by simultaneously solving the coupled set of nonlinear rate equations given in (A9) along with the wave equations in (4) and the Lorentzian equation in (5). Here, we utilize the FDTD algorithm to solve this system of equations along with proper boundary conditions. Details on the FDTD algorithm employed here are included in Appendix B. The boundary conditions for the system imply that the periodic waveguide is terminated at both sides by output waveguides (see Fig. 3) whose loading is represented by their characteristic impedances given in Appendix C (Table CIII). To study the operational scheme of the DBE lasers we first investigate the characteristics of the "cold" DBE cavity (also discussed in [47,76]), where for "cold" we mean absence of gain; however we pay attention to losses and include realistic parameters of the coupled waveguide.

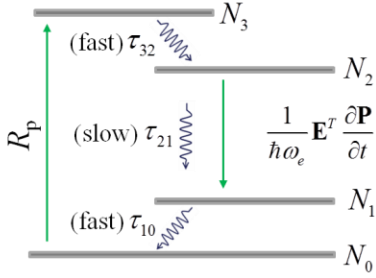

FIG. 4. A simplified four-level atomic system model in which lasing occurs between the atomic levels with population densities $N_2$ and $N_1$.

### C. Steady state gain medium response

In order to provide insight into characteristics of the DBE laser, the response of the "cold" DBE cavity is investigated as well as the steady state response of the DBE cavity with gain medium. Assuming time harmonic fields as $e^{-i\omega t}$, the electric field and polarization density amplitude vectors in (4) are given by $\mathbf{E}(z,t) = \mathrm{Re}\left[\mathbf{E}(z)e^{-i\omega t}\right]$ and $\mathbf{P}(z,t) = \mathrm{Re}\left[\mathbf{P}(z)e^{-i\omega t}\right]$, respectively, in which $\mathbf{E}(z)$ and $\mathbf{P}(z)$ are phasors.

In addition, we consider the active material population density at each energy level from the steady state point of view (at the steady state we have $dN_j/dt = 0$, $j = 0,1,2,3$). As such the population difference $\Delta N$ is constant and expressed as a function of the gain medium parameters; as typically done in steady state linearized gain models [84]. Therefore, the polarization density vector is simply related to the electric field vector through

$$\mathbf{P}(z) = \frac{\sigma_e}{\omega^2 - \omega_e^2 + i\omega\Delta\omega_e} \Delta N \, \underline{\underline{s}} \, \mathbf{E}(z) = \frac{g}{i\omega} \, \underline{\underline{s}} \, \mathbf{E}(z) \quad (7)$$

This equation defines the linearized gain parameter $g = i\omega\Delta N\sigma_e / \left(\omega^2 - \omega_e^2 + i\omega\Delta\omega_e\right)$ with unit of Siemens/m. This is analogous to the description where polarization density amplitudes are related to the electric field amplitudes through the susceptibility (see definition in page 494–541 in [83] and page 103-108 in [84]). The parameter $g$ represents gain/loss when its real part is negative/positive, whereas its imaginary part represents reactive loading due to the gain medium, under the small signal (linear) regime. In the example investigated here, only WG2 has active material (i.e., Erbium $Er^{3+}$) and therefore only the 2,2-entry, shown in Appendix C, of the gain coupling matrix $\underline{\underline{S}}$ is non-vanishing. Active medium parameters are detailed in Appendix C, and the plot of the gain parameter $g$, with unit of Siemens/m, is shown in Fig. 5(a) for a pumping rate of $R_p = 6 \times 10^6 \, \mathrm{s}^{-1}$. Fig. 5(a) shows that the gain parameter profile follows a Lorentzian shape with a negative real part within the frequency band of interest where, for simplicity, the periodic coupled waveguide in Fig. 3 is devised to have the DBE angular frequency $\omega_d$ coinciding with the emission angular frequency $\omega_e$.

### D. Cold DBE cavity characteristics

The properties of the "cold" DBE cavity, i.e., without the gain medium, are described here. We investigate the transfer function and the loaded $Q$-factor of the DBE cavity, using the transfer matrix analysis (refer to [47,65] for a thorough analysis of waveguides with DBE using the transfer matrix method). The analysis of cold DBE cavities has been done in various references, to mention a few [66,64,67,54,47], however here we only demonstrate the principal characteristics relative to lasing and the main contributing factors to lowering the lasing threshold and the single mode property in a DBE laser.

Note that the evolution equations of the wave amplitudes in the coupled waveguides can be described with first order differential coupled-wave equations that may be written in a Hermitian form (in the absence of gain and loss), as conventionally done in coupled-mode theory [86]. Therefore, this lossless system can be locally referred to as Hermitian (in the context of coupled-wave propagation [86,87]). In other words, the time harmonic coupled wave equations (1) are given in terms of the phasors $\mathbf{a}^{\pm}(z)$ with $\mathbf{a}^{\pm}(z,t) = \text{Re}\left[\mathbf{a}^{\pm}(z)e^{-i\omega t}\right]$ in a frequency domain description as such

$$\frac{\partial \mathbf{a}^{+}(z)}{\partial z} = ik_0 \underline{\underline{\mathbf{n}}} \, \mathbf{a}^{+}(z) \; ; \quad \frac{\partial \mathbf{a}^{-}(z,t)}{\partial z} = -ik_0 \underline{\underline{\mathbf{n}}} \, \mathbf{a}^{-}(z), \qquad (8)$$

where $k_0 = \omega/c$. By inspecting (8), one concludes that the matrix $k_0 \underline{\underline{\mathbf{n}}}$ is Hermitian since it is symmetric and therefore diagonalizable (refer to [44] for details) in each uniform segment of the lossless coupled waveguides.

The fundamental consequence of spatial periodicity of the lossless coupled waveguides under considerations (as those in Fig. 1) is that a non-uniform $\underline{\underline{\mathbf{n}}}(z)$ allows the EPD to occur even though the individual lossless uniform waveguides constitute Hermitian matrices as explained in detail in [88] using a transfer matrix formalism. In the following results are obtained using the TL formalism in Sec. II.B implemented via a transfer matrix formalism following [66,88], skipping the details here.

The finite length resonator is terminated by waveguides as shown in Fig. 3 and whose characteristic impedances are given in Appendix C (Table CIII). We first plot in Fig. 5(b) the transfer function [66].

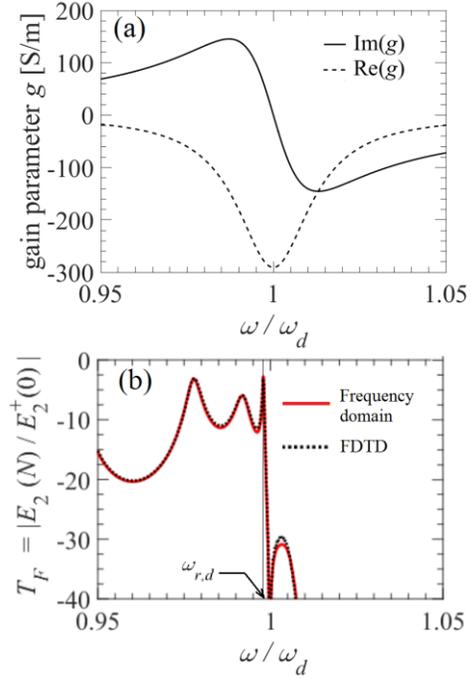

FIG. 5. (a) Active material steady state gain parameter $g$ that follows a Lorentzian line shape in frequency for a pump rate $R_p = 6 \times 10^6$ s$^{-1}$. (b) Transfer function of the cold DBE cavity, defined as $|E_2(N)/E_2^{+}(0)|$, with $N=20$, plotted versus frequency using both frequency domain analysis as well as the FDTD method explained in Appendix B. The DBE cavity resonance frequency (sharpest peak) occurs at $f_{r,d} \cong 193$THz (i.e., $\lambda_{r,d} \cong 1553$ nm).

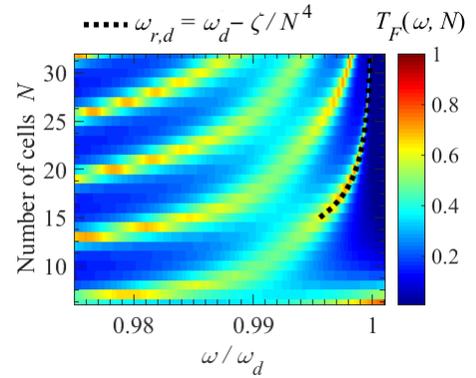

Fig. 6. Diagram of the transfer function of the cold DBE cavity versus the normalized frequency and the number of cells $N$. We also plot the fitting function $\omega_{r,d} = \omega_d - \zeta/N^4$, with $\zeta \approx 64\omega_d$ (dashed line) which shows how the DBE resonance (the sharpest resonance) approaches the DBE frequency.

Sharp transmission peaks near the DBE angular frequency $\omega_d$ are observed and the peak closest to the DBE frequency has the highest $Q$-factor [47], and we refer to it as the *DBE resonance* of the structure with finite length and we denote it by $\omega_{r,d}$. Note that such peak is the sharpest one, and in Fig. 5(b) occurs at $f_{r,d} = \omega_{r,d}/(2\pi) \cong 193$THz, i.e., at $\lambda_{r,d} = 2\pi c/\omega_{r,d} \cong 1553$ nm, and that several peaks are within the emission spectrum [Fig. 5(a)] of the gain material. The resonance frequency closest to the DBE frequency is dominant over all other resonances in the FPC with DBE. Because of the fourth power in the dispersion relation $(\omega_d - \omega) \approx h(k-k_d)^4$, a FPC resonance will occur at an angular frequency $\omega_{r,d}$ extremely close to $\omega_d$, where the group velocity vanishes, hence causing a very high density of states at $\omega_{r,d}$ [42]. This leads to the FPC resonance with highest group delay and highest local density of states (LDOS) as was shown in Ref. [47]. It is because of the largest density of states associated to the resonance peak at $\omega_{r,d}$, a single frequency operation of a laser cavity with DBE is expected, as it will be shown in Sec. IV.

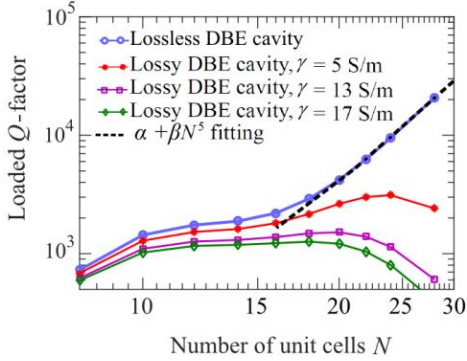

FIG. 7. Loaded $Q$-factor for "cold" DBE cavities with and without losses as a function of number of the cavity unit cells. We observe the unconventional scaling of the loaded $Q$-factor for "cold" DBE cavities as $N^5$ with $N$ being the number of unit cells. Here we show the calculated $Q$-factor as well as the fitting formula $\alpha+\beta N^5$ with $\alpha$ =767.7 and $\beta$ = 0.001. The quality of the fitting is provided by a root mean square error (RMSE) [64] of 0.998 for large $N$ ($N$>15). As shown in the figure, such unconventional scaling of $Q$ will be deteriorated by incorporating losses represented by increasing the distributed loss parameter $\gamma$.

For the sake of completeness, we compare the transfer function result obtained using the transfer matrix analysis with that calculated using the FDTD method that will be used later on in the paper. Fig. 5(b) shows identical agreement between both methods in analyzing the transfer function of the cold DBE cavity. In Fig. 6 we show the peaks of the transfer function varying number of unit cells in the periodic structure, calculated using the transfer matrix method. This figure shows that the transmission resonance frequencies are sharper when closer to the DBE frequency, and that the transmission resonance peak (i.e., the DBE resonance at $\omega_{r,d}$) shifts toward the DBE frequency $\omega_d$ as the number of cells $N$ increases, following the trajectories

$$\omega_{r,d} = \omega_d - \zeta/N^4, \quad (9)$$

where $\zeta = h(\pi/d)^4 \approx 64\omega_d$. This formula serves to estimate the working DBE resonance frequency as a function of number of unit cells. It should be observed that such DBE cavities have more or less twice the number of resonance frequencies as a uniform cavity with the same length, and that they are even closer to each other than those in a uniform cavity. Nevertheless, a single frequency of lasing operation will be demonstrated, due to the very high density of states of the resonance at $\omega_{r,d}$ [47].

In Fig. 7 we show some properties of the loaded $Q$-factor for DBE cavity. The loaded $Q$-factor is defined as $Q_{\text{tot}} = \omega_{r,d}(W_e + W_m)/P_L$ where $W_e$ and $W_m$ are the total electric and the magnetic time-average stored energies, while $P_L$ is the total time-average power loss. In the calculation of the power loss we consider the power dissipated in the material and that received by the loads. We plot the calculated loaded $Q$-factor at the DBE resonance versus the number of unit cells in Fig. 7 where we show how it scales with $N$. (Resonance is recalculated at each length, for each case.) In particular, we observe the unusual trend of the loaded $Q$-factor where the loaded $Q$-factor increases as $N^5$ for large $N$ for the lossless DBE cavity (as shown in Fig, 7 by the fitting formula: loaded $Q = a+bN^5$). However, dissipative losses in the dielectric [represented by the parameter $\gamma$ in (4) and in (A8)] limits such anomalous trend. In other words, the loaded $Q$-factor ceases to increase at one critical cavity length at which starts to deteriorate when losses overwhelm the response, i.e., eliminate transmission peak of the cavity. Note that for $\gamma > 17$ S/m, which is a high loss condition, the FPC with DBE composed of 20-unit cells would have negligible transmission and therefore the loaded $Q$-factor would be less than 1000. Such high loss cases are not considered here though may have advantages in certain class of lasers with PT-symmetry for instance [31,44]. Here, it is important to pay attention to the design of the DBE cavity by choosing the optimum number of cells to control the effect of losses and allow for the DBE resonance condition. We stress that the values of $Q$-factor (in Fig. 7) are only representative for the DBE cavity investigated here, and higher values of $Q$ could be obtained for an optimized design considering other implementations, utilizing CROWs for instances as in [50].

### III. DEGENERATE BAND EDGE (DBE) LASER

The evolution of the lasing dynamics in active DBE cavities is described by the nonlinear time-domain equations stated in Sec. II and Appendix A. In particular, the transient response of lasing action is well described by the evolution of gain in the cavity. In the small signal regime, photons within the DBE resonance at $\omega = \omega_{r,d}$ have the longest lifetime and the highest effective gain coefficient as defined in [42], among all other resonances in the cavity. As long as the overall gain experienced by such photons (electromagnetic waves) inside the cavity in a roundtrip is higher than the cavity losses (due to dissipative mechanism and escaping energy from the cavity ends), the intensity of the electromagnetic waves grows; and the DBE resonance has the highest growth rate among all other resonances in the cavity. Further analysis for the linear gain enhancement in DBE structures is established in [47].

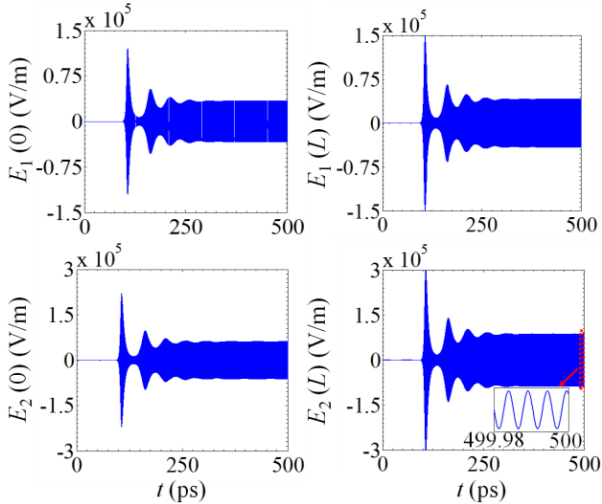

FIG. 8. Transient response of the electric field amplitudes recorded at all four waveguide outputs (Fig. 3) of the DBE laser (with $N=20$ unit cells) at $z=0$ and $z=L$ for a pumping rate of $R_\mathrm{p}=10^7\mathrm{s}^{-1}$. The zoomed area in right-bottom figure indicates the single frequency laser operation at the steady state. The laser steady state outputs are at a single frequency of ~192.5THz (~$\lambda$ = 1556.9 nm) which is very close to the DBE resonance frequency $f_{r,d}=193$ THz ($\lambda_{r,d}=1553$ nm).

On the other hand, for large electromagnetic wave intensities inside the cavity, nonlinearities are manifested in the rate equations, saturation occurs, and the output field amplitude reaches a steady-state. Here, we analyze the lasing action in the optical-waveguide-based DBE laser using the FDTD algorithm (see details in Appendix B). The parameters of the DBE laser are provided in Appendix C. Note that transient and steady state results are obtained here assuming that an initial short Gaussian pulse (whose parameters are provided in Appendix B) is launched into the WG2 from the left (Fig. 3).

The transient responses of the electric field amplitudes at all four waveguide outputs, for the case of a lossless DBE cavity, are plotted for a pumping rate of $R_\mathrm{p}=10^7\mathrm{s}^{-1}$ in Fig. 8. Note that the pumping rate of $10^7\mathrm{s}^{-1}$ used in Fig. 8 is larger than the lasing threshold pumping rate as we show later. We observe from Fig. 8 that the output field amplitudes saturate to the steady state at around $t_s \approx 350\,\mathrm{ps}$ thanks to the nonlinearity in the gain medium described in the nonlinear rate equations (see Appendix A). Note that the output electric field amplitudes in the waveguide outputs oscillate at a single frequency of ~192.5THz (~$\lambda$ = 1556.9 nm) as seen in Fig. 8 bottom, right panel; which is very close to the DBE cavity resonance frequency of $f_{r,d}$ ~193 THz (see Fig. 5), yet there

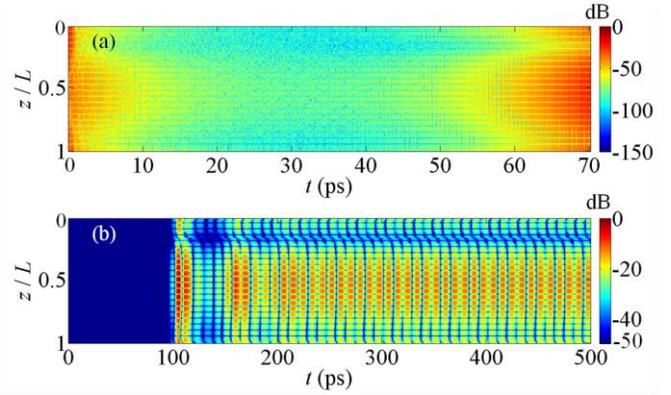

FIG. 9. The normalized electric field intensity (normalized to its maximum within the reported time window) inside the WG2 of the DBE laser cavity versus time: (a) 0-70 picoseconds interval, (b) 0-500 picoseconds interval. We observe a well-established oscillation with the DBE frequency after a short time.

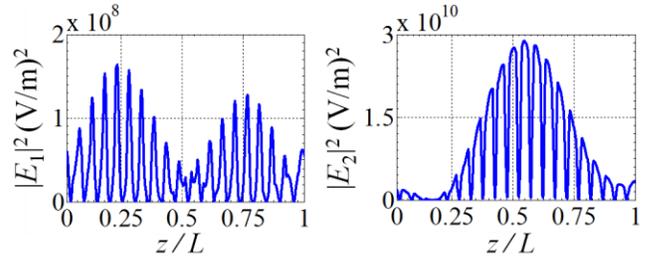

FIG. 10. Steady-state time-averaged electric field amplitude intensities inside the WG1 and the WG2 of the DBE laser cavity with $N$=20 ($L$=4.8μm). We observe that the steady-state field amplitudes profile resembles the resonance field amplitudes profile in the "cold" DBE cavity (i.e., without the gain medium) meaning that the DBE features are still occurring, even in the presence of non-linear gain.

is a very slight frequency shift due to the gain medium frequency pulling [83].

We first illustrate the mode selectivity by showing the space-time field evolution inside the cavity in Fig. 9 for which the DBE resonant field concentrated at the cavity center starts to grow exponentially after ~70 ps. Therefore; although many FPC resonances experience gain as seen in Fig. 5, the DBE resonance experiences the highest gain and dominates the output spectrum thanks to its unique field distribution and highest $Q$-factor. In addition, we plot the steady-state time-averaged electric field intensity inside the DBE laser cavity in Fig. 10. The steady-state field intensity is mainly concentrated near the cavity center for WG2; while the maximum field intensity in WG1 is at least two orders of magnitude lower than that in WG2. Nevertheless, the presence of WG1 is crucial to achieve the DBE. In addition, the steady-state field profile inside the DBE laser cavity resembles the field of the DBE resonance in the cold DBE cavity (not shown here for brevity, see Ref. [47,66]). Such observation indicates that the DBE features, which pertain to lossless and gainless periodic waveguides, are still persistent even when the cavity contains a non-linear gain. However, when operating just above the lasing threshold shown next, the main features of the DBE are retained. Very high levels of gain (i.e., very high pumping rates) may lead to other regimes of operation not considered in this paper since high gain may adversely deteriorate the DBE condition (see discussion in [47], and in [25] for RBE structures).

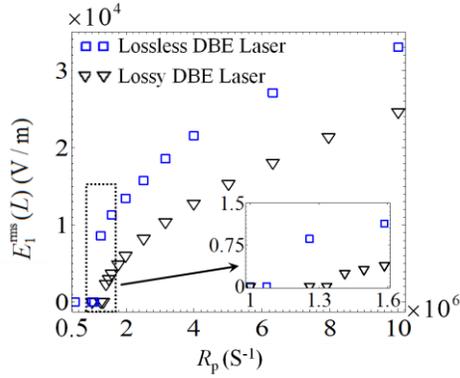

FIG. 11. The time average of the steady state electric field amplitude at the end of the WG1 (the root mean squared (RMS) defined as $E_1^{rms} = \sqrt{(T_2-T_1)^{-1} \int_{T_1}^{T_2} [E_1(t, z=L)]^2 dt}$ where the time interval T=$T_2-T_1$ is one period of the steady state electric field amplitude at the end of the WG1) of the lossless and lossy (with $\gamma = 13$ S/m ) DBE laser with $N=20$ unit cells, versus pumping rate. The comparative plot shows that the presence of the losses does not change the laser threshold dramatically. Moreover, the DBE laser maintains a single mode operation even when pumped up to 20 times of its threshold.

To calculate the lasing threshold for the DBE laser using time-domain simulations, we calculate the root mean square (RMS) value of the steady state electric field amplitude at the end of the WG1 of the cavity as a function of the pumping rate, and shown in Fig. 11. Two DBE laser cases with the 20 unit cells are investigated in Fig. 11, a lossless DBE laser with the total loaded $Q$-factor of 4200 and a lossy DBE laser with $\gamma = 13$ S/m (see Fig. 7) and a total loaded $Q$-factor of 3400. Remarkably, the DBE laser is shown to exhibit a single-frequency output even when pumped up to more than 20 times of its threshold as seen from Fig. 11. The threshold pumping rate $R_p^{th}$ is defined as the minimum pump rate that causes instability, i.e., the electric field amplitude to grow exponentially inside the cavity. Numerically, it is calculated as the value of $R_p$ at which the output field amplitude experiences a sudden transition from being around zero to having a significantly larger steady-state value. This is achieved by sweeping $R_p$ and observing the output steady state field amplitudes. As in Fig. 11, the root-mean-square (RMS) of the steady state output electric field amplitude (here recorded after 1.5 ns) for a lossless DBE laser is negligible (~0 V/m) for $R_p$ ~$1.08 \times 10^6$ s$^{-1}$ while it is significantly large (~$0.86 \times 10^4$ V/m) for $R_p$ ~$1.22 \times 10^6$ s$^{-1}$, indicating that a threshold pumping rate of approximately $R_p^{th}$ ~$1.15 \times 10^6$ s$^{-1}$ within a maximum error of ~6% due to finite numbers of simulation points. As such, the lasing pump threshold for the *lossless* and the *lossy* DBE lasers are $R_p^{th} \approx 1.15 \times 10^6$ s$^{-1}$ and $R_p^{th} \approx 1.4 \times 10^6$ s$^{-1}$ respectively. Furthermore, Fig. 11 shows that the RMS value of the steady state electric field amplitude of the laser output increases linearly with the pump rate above threshold.

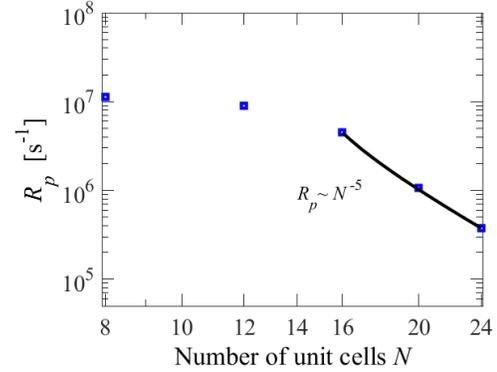

FIG. 12. Threshold pumping rate for the lossless DBE laser varying as a function of laser length (i.e., $Nd$) calculated using FDTD (squares), and plotting the fitting curve (for large number of unit cells) as a solid line with $R_p^{th} = R_0/(\alpha + \beta N^5)$, where $R_0=2\times 10^{10}$ s$^{-1}$, and the fitting constants $\alpha$ and $\beta$ are given in the caption of Fig. 7.

We also calculate, using the FDTD method, the threshold pumping rate for the lossless DBE laser with different numbers of unit cells following the same procedure we have used for calculating the lasing threshold for the DBE with $N=20$ unit cells in Fig. 11. Then, we plot the threshold pumping rate for the DBE laser varying as a function of the number of unit cells $N$ in Fig. 12. The trend of the threshold pumping rate for the DBE laser for large $N$ (i.e., $N \geq 16$) is also shown in Fig. 12 by the asymptotic fitting curve $R_p^{th} \propto N^{-5}$ which agrees with the theoretical calculations [47]. The reasons for such unconventional scaling of the threshold pumping rate with length is that loaded $Q$-factor of the DBE laser scales as $N^5$ (see [47,76] for more details), as compared to that of a conventional RBE laser which scales as $N^3$ [41], or that of a homogenous cavity that scales simply as $N$; as shown in Fig. 1(a), for the three regimes of operations discussed in this paper. We point out that for large number of unit cells DBE laser threshold becomes substantially lower by orders of magnitude than uniform FPC lasers as shown in Fig. 1(a). Interestingly, the constants of proportionality of the trend of $R_p^{th}$ as a function of $N$ are the same as those used to fit the loaded $Q$-factor in Fig. 7, indicating that $R_p^{th}(N) \sim Q^{-1}(N)$. It may seem apparent that the pumping threshold is only dictated by the loaded $Q$-factor, yet the field structure of the cavity plays a pivotal rule in lowering the threshold as we discuss in the following section.

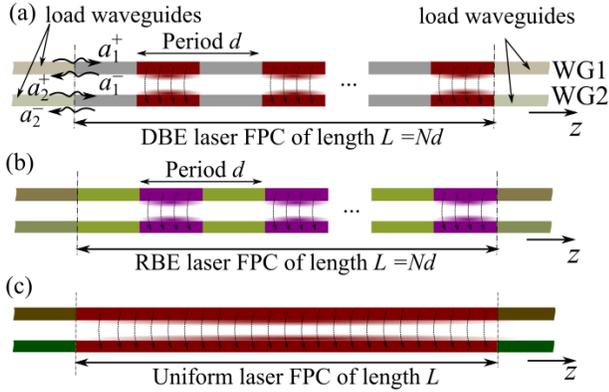

FIG. 13. Schematic geometries of the three FPC laser regimes compared in this paper: (a) FPC with DBE, (b) FPC with RBE, and (c) uniform FPC. Note that the FPC with DBE does not require mirrors at its ends, while they are essential in the uniform FPC, and somewhat necessary in the FPC with RBE, to ensure high $Q$-factor.

## IV. COMPARISON BETWEEN THE DBE LASER WITH OTHER CONVENTIONAL LASERS

To elaborate on the reasons of the superiority aspects of the DBE laser, we establish here a comparison between the proposed DBE laser with two other regimes of laser operations: (*i*) uniform FPC laser and (*ii*) RBE laser. The geometries of the three aforementioned laser cavities are shown in Fig. 13 where the DBE and RBE laser cases are corresponding to periodic structures operating near their band edge frequencies and comprising two periodically coupled waveguides, while the uniform FPC cavity is composed of two uniform (i.e., nonperiodic) coupled waveguides. Note that the choice of coupled waveguides for the RBE and the uniform cavity lasers is not necessary, however we here use two coupled-waveguides to preserve the analogy to the DBE laser for comparison purposes. Additionally, we aim at showing fundamentally unique and superior properties of DBE laser compared to the two other regimes. Therefore, the $Q$ factors and threshold values considered herein for all regimes of operation are shown as representative examples. Such unprecedented performance and the conclusions drawn thereafter are yet valid when implemented in other optical platforms supporting the DBE.

The dispersion relations and the magnitude of the transmission coefficients for the RBE and the DBE representative examples considered here are shown in Fig. 14(a)-(b), respectively. Indeed, we consider the aforementioned cold cavities (namely the FPCs with the DBE, the RBE, and the uniform case) such that they exhibit the same *resonance* frequency, length, and loaded $Q$-factor (that also takes into account losses as we will show later). This is achieved by adopting, for instance, the parameters for such cavities as in Appendix C. In order to achieve the same loaded $Q$-factor for the three aforementioned cavities with the same length, the reflectivity at the cavity ends is properly chosen as follows by resorting to the useful and general concept of waveguide impedance. We define a load power reflectivity at each waveguide (i.e., at each port) as the square of the respectively at $0.9985\omega_d$ and $0.99852\omega_d$. Note that the normalized field amplitude inside the DBE cavity is much stronger than that in the RBE cavity and is mainly concentrated in the middle of the cavity.

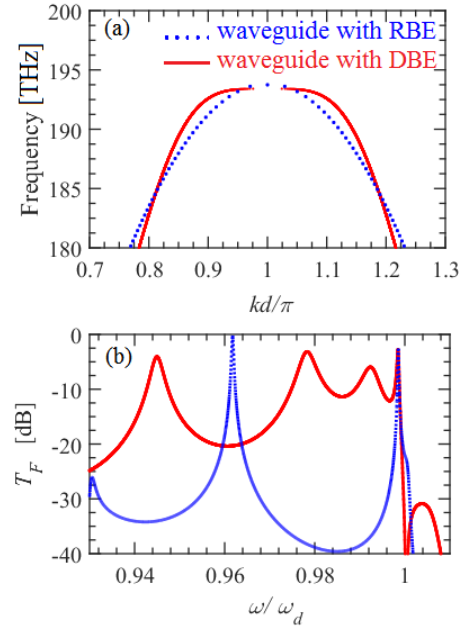

FIG. 14. (a) Dispersion relation of two "cold" periodic structures exhibiting, respectively, a DBE and a RBE, at $193.4 \, \text{THz}$ ($\lambda$=1550nm). (b) Magnitude of the transmission coefficients of the two finite length cavities with DBE and RBE that are developed to have the same resonance frequency (~193 THz), length, and loaded $Q$-factor ($Q = 4200$).

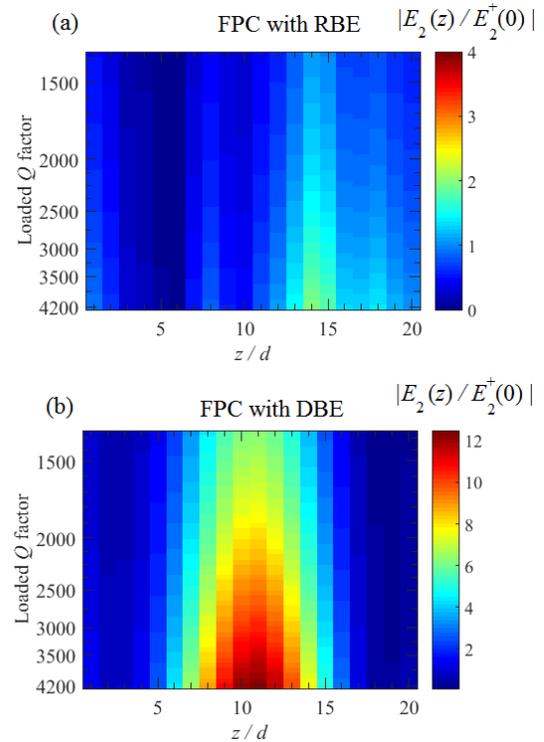

FIG. 15. Normalized electric field amplitude inside (a) the RBE "cold" FPC and (b) the DBE "cold" FPC varying as a function of the loaded $Q$-factor of the cavity. The loaded $Q$-factor decreases by incorporating dissipative losses. The field is plotted inside WG2 in both (a) and (b), which represents the maximum field enhancement for each case (when $Q$ = 4200) occurring

magnitude of the reflection coefficient at the interface between the output waveguide and the last FPC waveguide segment due to the difference in their characteristic impedances, as in Fig. 13 (see (C2) in Appendix C for definitions of reflectivity). Note that the output WGs' characteristic impedances for the three kinds of cavities as reported in Table CIII are varied to ensure that the three cavities have the same $Q$.

It is important to stress that the uniform cavity requires extremely high mirror reflectivity at each end (the power reflection coefficient is ~0.995, as defined in Appendix C) to have the same loaded $Q$-factor as the RBE and DBE cavities. Indeed, especially for the DBE case, reflection at the ends of the cavity is not realized by physical mirrors (i.e., large impedance mismatch between waveguide segments) since waveguide reflectivities are much lower. Indeed, a large reflection for the REB and especially for the DBE cavity is caused by mismatch between the degenerate Floquet-Bloch eigenwaves and the propagation eigenwaves in the external waveguides, and not by the simple characteristic impedance mismatch in each waveguide. Similar properties are also demonstrated in DBE CROWs [50] in which impedance mismatch or mirrors are not required for generating high $Q$-factor values.

We show in Fig. 15 the field distribution inside the "cold" coupled waveguides based FPCs operation at the resonance frequency for the two cases: the DBE cavity, considered in Sec. II-C and III, as well as the RBE cavity. The fields are plotted only for WG2 in both DBE and RBE FPCs and sampled only once in each unit cell along the cavity. The lossless cavities have a loaded $Q$-factor of ~4200 (the highest value of the $Q$-range considered here) and decreasing the loaded $Q$-factor occurs by introducing propagation losses in the waveguides using the parameter $\gamma$ in (4) and (A8). The large magnitude of the normalized field inside the DBE cavity demonstrates a uniquely-structured resonance field that leads to enhancing the local density of states [47] and therefore would lead to a lowered lasing threshold as compared to the RBE laser. On the contrary, the standing wave electric field magnitude inside a uniform FPC has a constant envelope and that envelope is independent on the resonant frequency which is not shown in Fig. 15 for brevity (see [30], [71] for details). It is worth mentioning that the field enhancement in the RBE cavity is not as high compared to the DBE counterpart (see Fig. 15). Indeed, the value of field enhancement in a cavity is markedly reliant on the topology, technology and length of the optical waveguides (for instance, a CROW with DBE reported in [50] can provide a higher field enhancement and higher $Q$-factors than the illustrative cases provided here). In essence, the maximum field enhancement in an FPC with DBE is at least 10 times that of an FPC with RBE having the same $Q$ factor and analogous topologies as seen in Fig. 15, and such observation can be generalized to other implementations of FPCs with DBE and RBE.

Next, we compare the three regimes of laser operation in terms of lasing threshold namely for the DBE, RBE and uniform cavities. Fig. 16 shows a comparison between the lasing threshold pump rate $R_p^{\text{th}}$ of the three aforementioned cavities as the loaded $Q$-factor varies. First and the foremost observation is that the lossless DBE laser (the maximum loaded $Q$-factor of ~4200 is given by termination loading) develops much lower threshold pump rate as compared to the lossless RBE laser of equal $Q$-factor and length; which in turn is also lower than that of the corresponding lossless uniform FPC laser. When losses are considered (loaded $Q$-factor less than 4200), the DBE laser has a lower threshold for all the ranges of the loaded $Q$-factor considered in Fig. 16 (i.e., from 1000 to 4200). For example, the DBE laser has 30% lower threshold than the uniform FPC laser of equal length when the loaded $Q$-factor is ~2000.

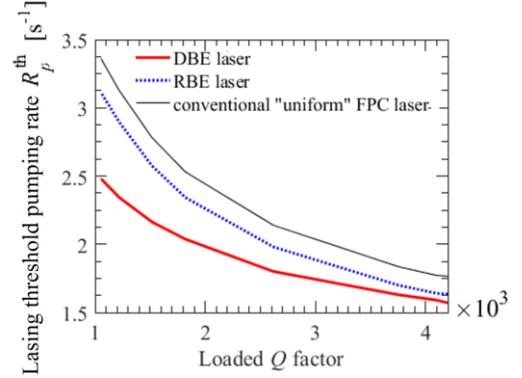

FIG. 16. Comparison between the lasing threshold pumping rate of the proposed DBE laser with the conventional uniform FPC and the RBE lasers of equal length ($L$=4.8μm) and resonant frequency (193 THz) varying as a function of their loaded $Q$-factor. The comparative plot clearly shows that the DBE laser has a significantly lower lasing threshold as compared to the RBE and the conventional uniform FPC lasers having the same $Q$-factor.

The results presented so far for the DBE laser assumed that the DBE cavity has mirrors with reflectivity at both ends, for simplicity of calculation, since for comparison we wanted to have (i) the same length, (ii) the same resonance frequency, and (iii) and the same $Q$-factor in all the three FPC types.

In general, the DBE resonant field confinement shown in Fig. 15(b) occurs even without mirror reflectors. Indeed, no mirrors are required for the DBE cavity to develop high $Q$-factor and strong field enhancement [47]. This is manifestly different from conventional uniform FPC laser cavities for which mirrors with high reflectivity are needed to reduce the lasing threshold [84]. We stress that in the DBE cavity, strong reflection of the DBE eigenwaves is still present, given by the degeneracy condition and almost unmatched Floquet-Bloch eigenwave impedance with the impedance of the load WG1 and WG2 waveguides, and not by the mismatch of the individual last waveguides segments to WG1 and WG2. The mirror reflectivity defined in Appendix C does not represent the Floquet-Bloch eigenwave reflection coefficient in the DBE case, rather it represents the reflection associated to each individual waveguide discontinuity, between the last waveguide segments of the FPC (from both right and left ends) and WG1 and WG2 outside the FPC. In other words, the high $Q$-factor of the DBE cavity relies on Floquet-Bloch eigenwave mismatch and not on mirror reflectors [21-22], [42]. (A Floquet-Bloch eigenwave characteristic impedance is represented by an impedance matrix [75]). In Fig. 17 we show the lasing pump rate threshold for the DBE laser varying as a function of its right-end and left-end mirror reflectivity (for this case they are assumed to be equal at each end, for both WG1 and WG2) compared to that for the uniform FPC laser having the same length. We observe that the DBE laser with no mirror reflectors (i.e., when the mirror power reflectivity is zero) has a pumping rate threshold which is an order of magnitude less that of the uniform FPC laser having ~93% mirror power reflection coefficient, as depicted from Fig. 17. This means that the DBE laser can be directly connected to the same waveguides WG1 and WG2 used in the FPC, i.e., without changing dimensions and without adding reflectors. Note that increasing the mirror reflectivity would dramatically decrease the lasing threshold for the uniform FPC laser, whereas the DBE laser has almost a steady threshold pump rate for reflectivity from 0 to 10%. As such, if we choose a mirror reflectivity of 98% for the uniform FPC laser; the corresponding uniform FPC length must substantially increase to achieve the same threshold of the DBE laser.

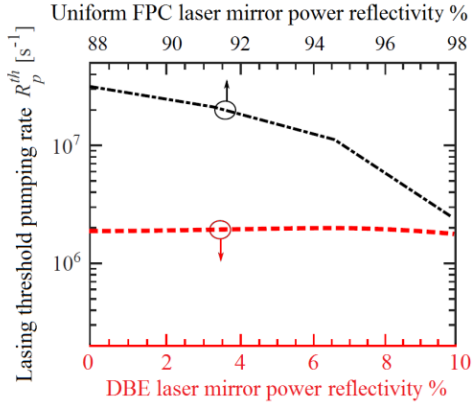

FIG. 17. Comparison between the lasing threshold pumping rate for the proposed DBE laser (dashed red) and the conventional uniform FPC (dash-dotted black) varying as a function of the mirror (load) power reflectivity. The DBE lasing threshold is significantly low regardless of the mirror reflectivity (even with no reflectivity at all) while the lasing threshold for the uniform FPC laser strongly depends on its mirror reflectivity [see definitions of reflectivity in Appendix C (C2)].

We finally show in Fig. 18 the excellent mode selection scheme of the DBE laser by observing the purity of its output field intensity spectrum in comparison with a *long* uniform FPC laser with relatively similar lasing threshold and same $Q$-factor; both operating at 1550 nm. The uniform FPC laser in this case is considered to have 98% mirror power reflection coefficient and also 10 times longer active medium length (i.e.,

$L$) than the DBE case in Sec. III (both are lossless cases for simplicity, and their parameters are provided in Appendix C). The load impedance of the DBE cavity ends is assumed to be the same as given in Appendix C (Table CIII). The length of the uniform FPC laser is chosen such that both the DBE and the uniform FPC lasers have an equal lasing threshold of the order of $10^6$ s$^{-1}$, and also a similar $Q$-factor of ~4000. Note that both cavities have multiple resonances within the gain emission spectrum.

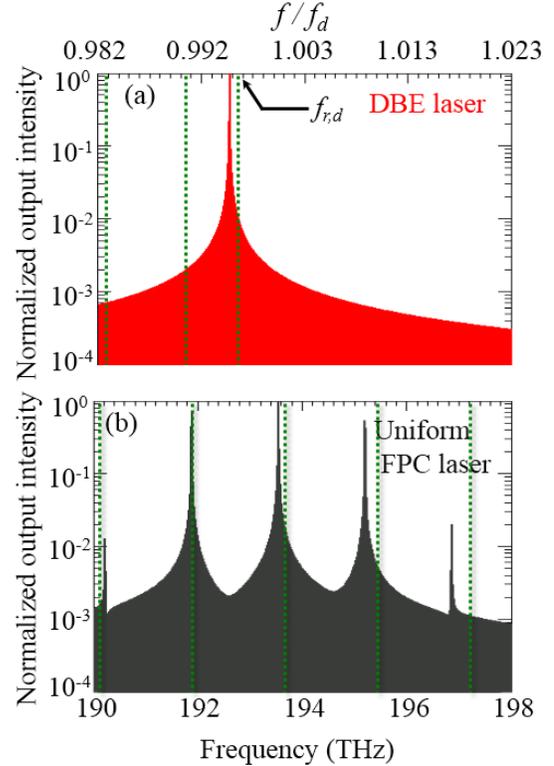

FIG. 18. Output spectrum of (a) the DBE laser and (b) the long uniform FPC laser at the same pumping rate of $R_p=10^7$ s$^{-1}$. The DBE laser in (a) features a single frequency operation (i.e., single oscillating mode) whereas the uniform FPC laser in (b) shows multiple frequencies of oscillation (i.e., multiple lasing modes) in the output spectrum. Vertical dashed lines denote the resonance frequencies of the cold FPCs, in both cases. For both plots, the spectrum is defined as the magnitude of the Fourier transform of the output field $E_1(t)$ taken in a time window from 0 to 1500 ps (for which steady state regime is obtained).

We plot in Fig. 18 the normalized output steady state field intensity spectrum for the DBE and uniform FPC lasers, both pumped above their respective threshold with $R_p=10^7$ s$^{-1}$ for both cases. The long "cold" uniform FPC has multiple resonances within the gain emission spectrum with a small free spectral range, i.e., spectral separation of the FPC resonance frequencies of ~1.6 THz. Moreover, such resonances have the same spectral width, therefore, they experience comparable gains, as discussed in [31,84]. This will in turn lead to multiple frequencies of oscillation within the laser (this is a typical scenario in conventional lasers, see page 42 in [30]). Indeed, the steady state output field intensity spectrum for the uniform FPC laser plotted in Fig. 18 shows multiple frequencies associated with the multiple lasing modes excited cavity (the resonant frequencies of the FPC without gain are depicted by dashed vertical lines in Fig. 18). On the contrary, the output intensity spectrum for the DBE laser seen in Fig. 18(a) shows a well-defined single-frequency (i.e., single oscillating mode) operation at ~192.5 THz near the cold DBE resonance frequency $f_{r,d}$ ~193 THz (as discussed earlier in Sec. II and Fig. 8). This indicates that the field of the DBE FPC resonance experiences a substantial *gain contrast* against all other FPC resonances (see Fig. 5), which leads to low-threshold lasing and mode selectivity when operating near the DBE. The remarkable single-frequency operation of the DBE low-threshold laser demonstrates its robustness as well as practicality for realizing single frequency coherent and low-noise sources.

## V. CONCLUSION

We have demonstrated a novel paradigm in light-matter interaction leading to low-threshold lasers based on exceptional point of degeneracy referred to as the degenerate band edge (DBE) in a pair of coupled periodic waveguides. We have provided the underlying theory behind such regime of operation of lasers that utilizes EPDs. In particular, we have demonstrated the DBE laser the features a significantly lower lasing threshold as compared to its conventional counterparts, i.e., a RBE laser and a uniform FPC laser, having the same resonant frequency, total length, and loaded $Q$-factor. The time-domain simulation results have also shown that the threshold pumping rate for the DBE laser exhibits an unprecedented scaling law of the threshold with length as $N^{-5}$, where $N$ is the number of unit cells. Non-linear gain media inclusion in the DBE cavity does not significantly perturb such mathematical condition, and the presented results have shown that the DBE laser can be pumped up to 20 times the threshold value and still maintain a single mode and a structured field distribution similar to that in the cold DBE cavity. The DBE laser does not require mirrors at its ends, indeed in some provided examples the mirrorless DBE cavity was terminated on a pair of waveguides identical to the ones inside the cavity. Furthermore, we have demonstrated the mode selection scheme associated with the DBE that leads to a single-frequency operation of the DBE laser in contrast to the conventional lasers that may operate with multiple oscillating frequencies. The novel paradigms proposed, and phenomenological conclusion drawn here, can be readily applied to provide robust threshold conditions, high efficiency and high output power, as well as low phase noise of not only optical but also infrared and terahertz sources.

## ACKNOWLEDGMENT

This material is based upon work supported by the Air Force Office of Scientific Research award number FA9550-15-1-0280 and under the Multidisciplinary University Research Initiative award number FA9550-12-1-0489 administered through the University of New Mexico. The authors also would like to thank CST of America, Inc. for providing CST.

## APPENDIX A: COUPLED WAVES FORMULATION AND RATE EQUATIONS

The evolution of the real-valued spatiotemporal wave amplitudes $\mathbf{a}^+(z,t)$ and $\mathbf{a}^-(z,t)$ are provided in (1)–(3) for a lossless and gainless uniform coupled waveguide. The system evolution matrix $\underline{\hat{\mathbf{M}}}$ in (3) is obtained from (2) and it is a 4×4 matrix given by

$$\underline{\hat{\mathbf{M}}} = \frac{1}{c}\begin{pmatrix} \underline{\underline{\mathbf{n}}} & \underline{\underline{\mathbf{0}}} \\ \underline{\underline{\mathbf{0}}} & -\underline{\underline{\mathbf{n}}} \end{pmatrix}. \tag{A1}$$

In this paper, it is convenient to resort to a fundamental description of waves propagating in the two coupled waveguides WG1 and WG2 (Figs. 1 and 3) using real-valued spatiotemporal field amplitudes $E_1(z,t)$, $H_1(z,t)$ and $E_2(z,t)$, $H_2(z,t)$, respectively. As discussed in Sec. II-B It is convenient to use the two-dimensional vectors $\mathbf{E}(z,t) = \begin{bmatrix} E_1(z,t) & E_2(z,t) \end{bmatrix}^T$ and $\mathbf{H}(z,t) = \begin{bmatrix} H_1(z,t) & H_2(z,t) \end{bmatrix}^T$. Note that wave amplitudes $\mathbf{a}^+$ and $\mathbf{a}^-$ are related to the total field amplitudes through a multiplication involving the characteristic impedance matrix as (see page 39 in [90] and also [74])

$$\mathbf{a}^+ = \frac{1}{2}\sqrt{\underline{\underline{\mathbf{Z}}}^{-1}}\left[\mathbf{E} + \underline{\underline{\mathbf{Z}}}\mathbf{H}\right], \quad \mathbf{a}^- = \frac{1}{2}\sqrt{\underline{\underline{\mathbf{Z}}}^{-1}}\left[\mathbf{E} - \underline{\underline{\mathbf{Z}}}\mathbf{H}\right] \tag{A2}$$

where $\underline{\underline{\mathbf{Z}}}$ is the 2×2 *characteristic impedance*

$$\underline{\underline{\mathbf{Z}}} = \begin{pmatrix} Z_{11} & Z_m \\ Z_m & Z_{22} \end{pmatrix}, \tag{A3}$$

symmetric and positive-definite matrix of the coupled waveguide (considered to piecewise uniform along $z$ and without gain or loss) which is defined using couple transmission line theory as in [78,91]. We recall that $\underline{\underline{\mathbf{Z}}}$ is defined as $\underline{\underline{\mathbf{Z}}} = \underline{\mathbf{T}}_E^{-1}\underline{\mathbf{T}}_H^{-1}$ where $\underline{\mathbf{T}}_E$ and $\underline{\mathbf{T}}_H$ are 2×2 matrices identified as the similarity transformations that bring the electric and magnetic field amplitude vectors, respectively, to their eigenwave, or *decoupled,* form (see details in page 94-100 in [91]). Therefore if we neglect dispersion (as discussed in Ch. 8 in [78] for instance) we can assume that $\underline{\underline{\mathbf{Z}}}$ is a real matrix and is used as a multiplier in (A2). Indeed, in a frequency domain description, we assume that the dispersion of $\underline{\underline{\mathbf{Z}}}$ and $\underline{\underline{\mathbf{n}}}$ is totally negligible compared to other dispersions in the system, i.e., those introduced by periodicity and the gain medium in the narrow frequency spectrum investigated here. Hence the refractive index and the characteristic impedance are purely real, and the normalization in (A2) is done through right-multiplying the spatiotemporal field amplitudes vectors with the square root and inverse of the impedance matrix $\underline{\underline{\mathbf{Z}}}$. Since the characteristic impedance matrix $\underline{\underline{\mathbf{Z}}}$ is positive-definite; the square roots taken in (A2) are defined as the unique positive-definite square root of the matrix $\underline{\underline{\mathbf{Z}}}$. Recall that a positive definite matrix has a unique positive definite square root which is obtained via diagonalization of the matrix and then by taking the principal square root (positive real part) of its real and positive eigenvalues, see Chapter 11 in [92].

It is then convenient to use the *total field amplitude state vector* $\mathbf{\Psi}(z,t)$ defined as $\mathbf{\Psi}(z,t) = \begin{bmatrix} \mathbf{E}^T(z,t) & \mathbf{H}^T(z,t) \end{bmatrix}^T$. Transformation (A2) allows to represent the state vector $\mathbf{\Psi}(z,t)$ in terms of the wave amplitude state vector $\hat{\mathbf{\Psi}}(z,t) = \begin{bmatrix} \left(\mathbf{a}^+(z,t)\right)^T & \left(\mathbf{a}^-(z,t)\right)^T \end{bmatrix}^T$ in (3) by a simple transformation

$$\mathbf{\Psi} = \underline{\mathbf{U}}\,\hat{\mathbf{\Psi}}\ . \tag{A4}$$

The 4×4 matrix $\underline{\mathbf{U}}$ is a transformation of the physical electric and magnetic field amplitudes to the normalized wave (scattering) amplitudes which follows from (A2) and given by

$$\underline{\mathbf{U}} = \begin{bmatrix} \sqrt{\underline{\underline{\mathbf{Z}}}} & \sqrt{\underline{\underline{\mathbf{Z}}}} \\ \sqrt{\underline{\underline{\mathbf{Z}}}^{-1}} & -\sqrt{\underline{\underline{\mathbf{Z}}}^{-1}} \end{bmatrix}. \tag{A5}$$

The corresponding system equation for the field amplitudes state vector $\mathbf{\Psi}(z,t)$ is given by

$$\frac{\partial \mathbf{\Psi}(z,t)}{\partial z} = -\underline{\mathbf{M}}\frac{\partial \mathbf{\Psi}(z,t)}{\partial t}, \tag{A6}$$

Using the transformation (A3) between $\mathbf{\Psi}(z,t)$ and $\hat{\mathbf{\Psi}}(z,t)$ we thereby find the system matrix $\underline{\mathbf{M}}$ in (A2) and (4) in terms of the other system matrix $\hat{\underline{\mathbf{M}}}$ in (3) as

$$\underline{\mathbf{M}} = \underline{\mathbf{U}}\,\hat{\underline{\mathbf{M}}}\,\underline{\mathbf{U}}^{-1} = \frac{1}{c}\begin{pmatrix} \underline{\mathbf{0}} & \sqrt{\underline{\underline{\mathbf{Z}}}}\,\underline{\underline{\mathbf{n}}}\sqrt{\underline{\underline{\mathbf{Z}}}} \\ \sqrt{\underline{\underline{\mathbf{Z}}}^{-1}}\,\underline{\underline{\mathbf{n}}}\sqrt{\underline{\underline{\mathbf{Z}}}^{-1}} & \underline{\mathbf{0}} \end{pmatrix}. \tag{A7}$$

Further discussion about such transformations are found in [93]. In general, matrices $\underline{\underline{\mathbf{Z}}}$ and $\underline{\underline{\mathbf{n}}}$ do not necessarily commute. However, based on our choice of parameters for all the cases provided in this paper, matrices $\underline{\underline{\mathbf{n}}}$ and $\underline{\underline{\mathbf{Z}}}$ commute (i.e., $\underline{\underline{\mathbf{n}}}\underline{\underline{\mathbf{Z}}} = \underline{\underline{\mathbf{Z}}}\underline{\underline{\mathbf{n}}}$), which also means that $\sqrt{\underline{\underline{\mathbf{Z}}}}\,\underline{\underline{\mathbf{n}}}\sqrt{\underline{\underline{\mathbf{Z}}}} = \underline{\underline{\mathbf{n}}}\underline{\underline{\mathbf{Z}}}$. Indeed, the reason of why matrices $\underline{\underline{\mathbf{n}}}$ and $\underline{\underline{\mathbf{Z}}}$ commute is because the two coupled waveguides in each segment of the periodic cell are identical, as shown in Appendix C. In other words, the diagonal entries of matrixes $\underline{\underline{\mathbf{Z}}}$ and $\underline{\underline{\mathbf{n}}}$ for the coupled segments of the waveguide are equal; therefore $\underline{\underline{\mathbf{Z}}}$ and $\underline{\underline{\mathbf{n}}}$ commute. For the uncoupled segments, $\underline{\underline{\mathbf{Z}}}$ and $\underline{\underline{\mathbf{n}}}$ are diagonal matrices therefore they commute. The coupled waveguide analysis in this paper can be also represented using conventional coupled-mode theory [69–74], however this is outside the scope of this paper, and as mentioned in Sec. II-B we found convenient to use the time domain transmission line formulation in (4) that is readily implemented in the FDTD [79] [80–82].

The load waveguides on the left and right sides of the cavity are assumed not coupled to each other and they are also modeled using their characteristic impedances, which can be cast as the diagonal elements of a diagonal matrix $\underline{\underline{\mathbf{Z}}}_L$. The load waveguides at the right and left ends of the cavity can be different from each other.

Gain and losses in the system are included via the polarization amplitude $\mathbf{P}$ and $\underline{\underline{\boldsymbol{\gamma}}}$ in (4) that describes the TD evolution, and not via the $\hat{\underline{\mathbf{M}}}$ and $\underline{\mathbf{M}}$ matrices in (3) and (4). The per unit length loss parameter $\underline{\underline{\boldsymbol{\gamma}}}$ in (4) is a 2×2 matrix given by

$$\underline{\underline{\boldsymbol{\gamma}}} = \begin{pmatrix} \gamma & 0 \\ 0 & \gamma \end{pmatrix}, \tag{A8}$$

where $\gamma$ has the unit of Siemens/m and represents per-unit-length loss in the coupled waveguides. In a transmission line formalism, each parameter $\gamma$ represents the per-unit-length shunt conductance. The gain media is represented by a four-level energy system as schematically shown in Fig. 4. The dynamics of the population densities of different energy levels are dictated by the nonlinear rate equations given below. In this regard, the population density of each level denoted as $N_j$ (with $j = 0,1,2,3$) is space-and time dependent and given by [82,84,94]

$$\frac{\partial N_3(z,t)}{\partial t} = N_0(z,t)R_p - \frac{N_3(z,t)}{\tau_{32}},$$

$$\frac{\partial N_2(z,t)}{\partial t} = \frac{N_3(z,t)}{\tau_{32}} + \frac{1}{\hbar\omega_e}\left[\underline{\mathbf{s}}\mathbf{E}(z,t)\right]^T \frac{\partial \mathbf{P}(z,t)}{\partial t} - \frac{N_2(z,t)}{\tau_{21}},$$

$$\frac{\partial N_1(z,t)}{\partial t} = \frac{N_2(z,t)}{\tau_{21}} - \frac{1}{\hbar\omega_e}\left[\underline{\mathbf{s}}\mathbf{E}(z,t)\right]^T \frac{\partial \mathbf{P}(z,t)}{\partial t} - \frac{N_1(z,t)}{\tau_{10}},$$

$$\frac{\partial N_0(z,t)}{\partial t} = \frac{N_1(z,t)}{\tau_{10}} - N_0(z,t)R_p, \tag{A9}$$

where $T$ denotes matrix transpose and $\left[\underline{\underline{\mathbf{s}}}\mathbf{E}(z,t)\right]^T \partial \mathbf{P}(z,t)/\partial t = \sum_{\mu\nu} s_{\mu\nu} E_\nu \, \partial P_\mu / \partial t$ and $\hbar$ is the reduced Planck's constant. The sum of the population densities is also equal to the total active material doping density, $N_T = \sum_{j=0}^{3} N_j$. The waveguide and gain medium physical parameters used in the examples in this paper are given in Appendix C.

## APPENDIX B: FDTD ALGORITHM FOR COUPLED WAVEGUIDE CAVITY WITH GAIN

In this Appendix, we describe the FDTD algorithm used in this paper for the analysis of the coupled waveguide interacting with gain media featuring the space-time evolution of the waveguide electric and magnetic fields' amplitudes $\mathbf{E}(z,t) = [E_1(z,t) \quad E_2(z,t)]^T$ and $\mathbf{H}(z,t) = [H_1(z,t) \quad H_2(z,t)]^T$. We discretize the computational domain (time and space) based on the Yee algorithm [79,80], in one dimension, such that the electric field amplitude is stored at integer node positions (in the $z$ direction) while staggered by $\Delta t/2$ in time, namely $\mathbf{E}_i^{n+1/2} = \mathbf{E}(i\Delta z, [n+1/2]\Delta t)$. Here, $i$ and $n$ are integers and $\Delta z$ and $\Delta t$ are grid intervals in space and time, respectively. The magnetic field amplitude on the other hand is stored at integer times while staggered by $\Delta z/2$ in the $z$ direction, i.e., $\mathbf{H}_{i+1/2}^{n+1} = \mathbf{H}([i+1/2]\Delta z, (n+1)\Delta t)$. The population density and polarization density amplitude vectors are also sampled at the same locations and times as the electric field amplitude. Using the central difference approximation, the time dependent differential equations in (4) and (5) are, respectively, written in the discrete form as

$$\begin{aligned}
\mathbf{E}_i^{n+1/2} &= -c\Delta t \left(\underline{\underline{\mathbf{n}}}_i^{-1} \underline{\underline{\mathbf{Z}}}_i\right)\left(\mathbf{H}_{i+1/2}^n - \mathbf{H}_{i-1/2}^n\right)/\Delta z \\
&\quad - c\left(\underline{\underline{\mathbf{n}}}_i^{-1} \underline{\underline{\mathbf{Z}}}_i\right)\left(\mathbf{P}_i^{n+1/2} - \mathbf{P}_i^{n-1/2}\right) - \\
&\quad c\Delta t \left(\underline{\underline{\mathbf{n}}}_i^{-1} \underline{\underline{\mathbf{Z}}}_i\right) \underline{\underline{\boldsymbol{\gamma}}}_i \left(\mathbf{E}_i^{n+1/2} + \mathbf{E}_i^{n-1/2}\right)/2 + \mathbf{E}_i^{n-1/2},
\end{aligned}$$ (B1)

$$\mathbf{H}_{i+1/2}^{n+1} = -c\Delta t \left(\underline{\underline{\mathbf{n}}}_i \underline{\underline{\mathbf{Z}}}_i\right)^{-1}\left(\mathbf{E}_{i+1}^{n+1/2} - \mathbf{E}_i^{n+1/2}\right)/\Delta z + \mathbf{H}_{i+1/2}^n,$$

and

$$\begin{aligned}
\mathbf{P}_i^{n+1/2} &= \frac{2\Delta t^2}{2+\Delta\omega_e \Delta t}\left[-\sigma_e \Delta N_i^{n-1/2} \underline{\underline{\mathbf{s}}}\mathbf{E}_i^{n-1/2} + \right. \\
&\quad \left. + \left(2/\Delta t^2 - \omega_e^2\right)\mathbf{P}_i^{n-1/2} + \left(\Delta\omega_e\Delta t - 2\right)/\left(2\Delta t^2\right)\mathbf{P}_i^{n-3/2}\right],
\end{aligned}$$ (B2)

Accordingly, the nonlinear rate equations for the four-level atomic system [given in (A9)] are also discretized as

$$\begin{aligned}
N_{3,i}^{n+1/2} &= B_3 \left[2R_p N_{0,i}^{n-1/2} + A_3 N_{3,i}^{n-1/2}\right] \\
N_{2,i}^{n+1/2} &= B_2 \left[\left(N_{3,i}^{n+1/2} + N_{3,i}^{n-1/2}\right)/\tau_{32} + A_2 N_{2,i}^{n-1/2} + \right. \\
&\quad \left. \left(\mathbf{E}_i^{n+1/2} + \mathbf{E}_i^{n-1/2}\right)^T \left(\mathbf{P}_i^{n+1/2} - \mathbf{P}_i^{n-1/2}\right)/\left(\hbar\omega_e \Delta t\right)\right] \\
N_{1,i}^{n+1/2} &= B_1 \left[\left(N_{2,i}^{n+1/2} + N_{2,i}^{n-1/2}\right)/\tau_{21} + A_1 N_{1,i}^{n-1/2} - \right. \\
&\quad \left. \left(\mathbf{E}_i^{n+1/2} + \mathbf{E}_i^{n-1/2}\right)^T \left(\mathbf{P}_i^{n+1/2} - \mathbf{P}_i^{n-1/2}\right)/\left(\hbar\omega_e \Delta t\right)\right] \\
N_{0,i}^{n+1/2} &= \varsigma \left[\left(N_{1,i}^{n+1/2} + N_{1,i}^{n-1/2}\right)/\tau_{10} + N_{0,i}^{n-1/2}\left(2 - \Delta t R_p\right)/\Delta t\right]
\end{aligned}$$ (B3)

where

$$\varsigma = \frac{\Delta t}{2+\Delta t R_p}; \quad A_m = \left(2\tau_{m,m-1} - \Delta t\right)/\left(\Delta t \tau_{m,m-1}\right);$$

$$B_m = \frac{\tau_{m,m-1}\Delta t}{2\tau_{m,m-1} + \Delta t}; \quad m = 1,2,3$$ (B4)

Explicit equations (B1)-(B3) compose the complete FDTD update equations for a coupled waveguide with an optically pumped four-level gain medium. The computation is a three-step recursive process: (*i*) the polarization density amplitude vector is calculated through (B2), (*ii*) the electric and magnetic field amplitudes are calculated using (B1), and (*iii*) the population density at each energy level is then calculated from (B3). This three-step recursive process is repeated until the end of the simulation time.

## APPENDIX C: PARAMETERS OF COUPLED WAVEGUIDES AND GAIN MEDIA

The periodic waveguide has been developed to have a DBE at frequency $f_d = \omega_d/(2\pi) = 193.5$ THz (i.e., free-space wavelength $\lambda_d = 1550$ nm), which coincides with the dipolar emission frequency of the active material in WG2 $f_e = 193.5$ THz. The other periodic coupled waveguide used for comparison throughout the paper is developed to have an RBE at the same frequency. Throughout the paper, we have assumed that the gain media is present only in WG2 and only the field in WG2 interacts with the gain medium. Therefore, the matrix $\underline{\underline{\mathbf{s}}}$ in (4) and (7) is taken as

$$\underline{\underline{\mathbf{s}}} = \begin{pmatrix} 0 & 0 \\ 0 & 1 \end{pmatrix}.$$ (C1)

The unit cell in the periodic structure in Figs. 1, 3 and 13 is made of two segments. The characteristic impedance matrix $\underline{\underline{\mathbf{Z}}}$ and refractive index matrix $\underline{\underline{\mathbf{n}}}$ of the first (uncoupled) segment, and the second (coupled) segment (see Figs. 1 and 13) of the constitutive lossless coupled waveguides with DBE and RBE are, respectively, given in Tables CI and CII. These parameters can be realized using Silicon ridge waveguides [95,96] using the geometry in Fig. C1. The parameters for the uniform coupled waveguide used for comparison are also provided in Tables CI and CII.

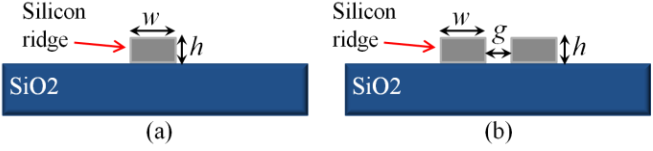

FIG. C1. Cross section of (a) uncoupled (single), and (b) coupled waveguides with two identical rectangular Silicon ridges for which the parameters in Tables C1-CIII are based on.

As an illustrative example, just to prove the proposed DBE laser concepts, the lengths of the uncoupled and the coupled sections (see Fig. 1) are $d_1/2 = 2d_2 = 0.096\,\mu\text{m}$. The DBE and RBE laser cavities are here composed of 20 unit cells with period of $d = d_1 + d_2 = 0.24\,\mu\text{m}$ and have a total length of $L = 4.8\,\mu\text{m}$. In our examples, we have taken the parameters in Tables CI-CIII to model Silicon ridge waveguides on a SiO2 substrate (as depicted schematically in Fig. 1(a) as well as in Ref. [78]). The cross-section of each uncoupled waveguide is shown in Fig. C1(a) and is composed of a Silicon rectangular ridge of width $w$ and height $h$, while the coupled waveguides' cross-section is shown in Fig. C1(b) and is composed of two identical Silicon rectangular ridges in proximity, with lateral gap spacing of $g$. For example, in the DBE coupled waveguide geometry, the two uncoupled waveguides are identical with $w = 520$ nm and $h = 400$ nm; moreover, the coupled waveguides are also identical and their parameters are $w = 800$ nm, $h = 400$ and $g = 250$ nm. The corresponding parameters in Tables C1-CIII are then obtained based on full-wave simulations of the coupled Silicon ridge waveguides, from which the effective refractive indices as well as the characteristic impedances are extracted (full-wave simulations were carried out using CST Microwave Studio, frequency domain solver based on the finite element method).

The left and right mirror power reflectivities associated to the loading waveguides at the two ends of the cavity (Fig. 13) are calculated as follows. We consider the impedances of the loads of WG1 and WG2 (i.e., $Z_{L1}^r$ and $Z_{L2}^r$ at the right end of the cavity, and $Z_{L1}^l$ and $Z_{L2}^l$ at the left end of the cavity) and the characteristic impedance entries of the first and last segments of the two waveguides upon which the cavity is terminated (i.e., $Z_{11}^c$ and $Z_{22}^c$ for the coupled segment and i.e., $Z_{11}^u$ and $Z_{22}^u$ for the uncoupled one). The superscripts $c$ and $u$ denote the coupled and uncoupled segments respectively, while $r$ and $l$ denote right and left ends of the cavity, respectively. Accordingly, we define the four mirror power reflectivity terms as

$$R_1^r = \left|\frac{Z_{L1}^r - Z_{11}^c}{Z_{L1}^r + Z_{11}^c}\right|^2, \quad R_2^r = \left|\frac{Z_{L2}^r - Z_{22}^c}{Z_{L2}^r + Z_{22}^c}\right|^2$$
$$R_1^l = \left|\frac{Z_{L1}^l - Z_{11}^u}{Z_{L1}^l + Z_{11}^u}\right|^2, \quad R_2^l = \left|\frac{Z_{L2}^l - Z_{22}^u}{Z_{L2}^l + Z_{22}^u}\right|^2 \quad (C2)$$

where $R_1^r$ and $R_2^r$ are the power reflectivities of the WG1 and WG2 respectively, at the right end of the cavity, whereas $R_1^l$ and $R_2^l$ are the power reflectivities of the WG1 and WG2 respectively, at the left end of the cavity.

In all the cases reported in this paper (except for those in Fig. 17), the characteristic impedances of the coupled and uncoupled waveguides and load impedances for the DBE, RBE, and uniform FPCs (see Fig. 13) are reported in Tables CI through CIII. Therefore, the power reflectivities can be readily calculated from (C2) for each laser cavity. Moreover, the load waveguides of WG1 and WG2 are different from each other; while the right and left load impedances are equal for WG1 as well as for WG2 (in other words $Z_{L1}^{r,l} \neq Z_{L2}^{r,l}$ while $Z_{L1}^r = Z_{L1}^l$ and $Z_{L2}^r = Z_{L2}^l$) and such differences are used to tailor the Q-factors and the thresholds of the three types of laser cavities and for comparison purposes. Note that the uniform FPC is symmetric from both right and left ends therefore we have $R_1^l = R_1^r$ and $R_2^r = R_2^l$.

Only for the results presented in Fig. 17 the mirror reflectivity is varied in the DBE cavity, as well as for the uniform laser cavity, contrary to those in Table CIII. To produce the results in Fig. 17, load impedances for WG1 and WG2, for both left and right ends of the cavities, are chosen such that the reflectivity of all four mirrors is the same. In other words, for Fig. 17 we choose $R_1^r = R_1^l = R_2^r = R_2^l$, for both cases of the DBE and the uniform FPC laser cavities which is then varied and the corresponding lasing threshold is plotted for both cases in Fig. 17. Note that the mirror reflectivity values considered in this paper can be readily realized using Silicon ridge waveguides [95,96].

TABLE CI. The self and mutual effective refractive indices of the coupled and uncoupled sections of the waveguide forming the DBE and the RBE. We report also the coupled waveguides parameters for the uniform cavity.

|         | Uncoupled section | | Coupled section | |
|---------|-------------------|-----|-----------------|-----|
|         | $n_{11}$ | $n_{22}$ | $n_{11}=n_{22}$ | $n_m$ |
| DBE     | 2.81 | 2.81 | 2.03 | 1.32 |
| RBE     | 2.2  | 2.98 | 2.47 | 0.49 |
| Uniform | -    | -    | 2.6  | 1.65 |

TABLE CII. The self and mutual characteristic impedances for the coupled and uncoupled sections of the coupled waveguide with DBE and RBE, and for the uniform waveguide.

| Uncoupled section | Coupled section |
|-------------------|-----------------|

|  | $Z_{11}^u$ (Ω) | $Z_{22}^u$ (Ω) | $Z_{11}^c = Z_{22}^c$ (Ω) | $Z_m^c$ (Ω) |
|---|---|---|---|---|
| DBE | 159 | 108 | 272 | 187 |
| RBE | 159 | 108 | 155 | 44 |
| Uniform | - | - | 272 | 187 |

TABLE CIII. Characteristic impedances of the output WG1 and WG2 coupled to the laser cavities with DBE, RBE and uniform one; both right and left ends of the cavity are assumed to be have equal load impedances in all cases of the paper, except for the results in Fig. 17.

| Load waveguides impedance | |
|---|---|
| $Z_{L1}^l = Z_{L1}^r$ (kΩ) | $Z_{L2}^l = Z_{L2}^r$ (kΩ) |
| DBE  5 | 3 |
| RBE  12 | 8 |
| Uniform  375 | 260 |

The gain is provided by having optically-pumped active atoms (e.g., here $Er^{3+}$ as described in [70]) doped in the WG2 cavity segment of the coupled waveguide. The photon lifetime of the transitions between the energy levels in the $Er^{3+}$ (see Fig. 4) are $\tau_{10} = 0.1\,\mathrm{ps}$, $\tau_{21} = 300\,\mathrm{ps}$ and $\tau_{32} = 0.1\,\mathrm{ps}$ [51]. The emission frequency and gain bandwidth are also $f_e = 193.5\,\mathrm{THz}$ and $\Delta f_e = 5\,\mathrm{THz}$, respectively, and the initial ground-state electron density in the WG2 material is $N_0 = 5 \times 10^{23}\,\mathrm{m}^{-3}$ [90]. These gain medium parameters are assumed to be constant and independent of lasing process.

In our FDTD simulations, we choose the space discretization step of $\Delta z = 5.99\,\mathrm{nm}$ and the time discretization step is set to be $\Delta t = 3.176 \times 10^{-6}\,\mathrm{ps}$, which is sufficiently small to have a numerically stable FDTD algorithm. The oscillation process is also initiated by launching a short Gaussian pulse of peak amplitude 1V/m, full width at half maximum (FWHM) of $\Delta t \times 10^4$ (where $\Delta t$ is the FDTD grid interval in time), peak's time of $1.3 \times$ FWHM and modulated at the DBE wavelength ($\lambda$=1550nm), into the cavity.

## REFERENCES

yaz[1] E. Yablonovitch, Phys. Rev. Lett. **58**, 2059 (1987).
[2] M. Lončar, T. Yoshie, A. Scherer, P. Gogna, and Y. Qiu, Appl. Phys. Lett. **81**, 2680 (2002).
[3] M. Meier, A. Mekis, A. Dodabalapur, A. Timko, R. E. Slusher, J. D. Joannopoulos, and O. Nalamasu, Appl. Phys. Lett. **74**, 7 (1999).
[4] S.-L. Chua, Y. Chong, A. D. Stone, M. Soljacic, and J. Bravo-Abad, Opt. Express **19**, 1539 (2011).
[5] S. Noda, M. Yokoyama, M. Imada, A. Chutinan, and M. Mochizuki, Science **293**, 1123 (2001).
[6] K. Srinivasan, P. E. Barclay, O. Painter, J. Chen, A. Y. Cho, and C. Gmachl, Appl. Phys. Lett. **83**, 1915 (2003).
[7] T. Yoshie, J. Vučković, A. Scherer, H. Chen, and D. Deppe, Appl. Phys. Lett. **79**, 4289 (2001).
[8] H. Altug and J. Vuckovic, Opt. Express **13**, 8819 (2005).
[9] B.-S. Song, S. Noda, T. Asano, and Y. Akahane, Nat. Mater. **4**, 207 (2005).
[10] Y. Akahane, T. Asano, B.-S. Song, and S. Noda, Opt. Express **13**, 1202 (2005).
[11] Q. Quan and M. Loncar, Opt. Express **19**, 18529 (2011).
[12] Y. Akahane, T. Asano, B.-S. Song, and S. Noda, Nature **425**, 944 (2003).
[13] M. H. Shih, W. Kuang, A. Mock, M. Bagheri, E. H. Hwang, J. D. O'Brien, and P. D. Dapkus, Appl. Phys. Lett. **89**, 101104 (2006).
[14] T. Tanabe, M. Notomi, E. Kuramochi, A. Shinya, and H. Taniyama, Nat. Photonics **1**, 49 (2007).
[15] M. Notomi, E. Kuramochi, and H. Taniyama, Opt. Express **16**, 11095 (2008).
[16] W. Zhou, M. Dridi, J. Y. Suh, C. H. Kim, D. T. Co, M. R. Wasielewski, G. C. Schatz, and T. W. Odom, Nat. Nanotechnol. **8**, 506 (2013).
[17] T. K. Hakala, H. T. Rekola, A. I. Väkeväinen, J.-P. Martikainen, M. Nečada, A. J. Moilanen, and P. Törmä, Nat. Commun. **8**, 13687 (2017).
[18] A. H. Schokker, F. van Riggelen, Y. Hadad, A. Alù, and A. F. Koenderink, Phys. Rev. B **95**, 085409 (2017).
[19] M. Nomura, S. Iwamoto, A. Tandaechanurat, Y. Ota, N. Kumagai, and Y. Arakawa, Opt. Express **17**, 640 (2009).
[20] F. Raineri, A. M. Yacomotti, T. J. Karle, R. Hostein, R. Braive, A. Beveratos, I. Sagnes, and R. Raj, Opt. Express **17**, 3165 (2009).
[21] A. Figotin and I. Vitebskiy, Laser Photonics Rev. **5**, 201 (2011).
[22] A. Figotin and I. Vitebskiy, Phys. Rev. E **72**, 036619 (2005).
[23] A. Figotin and I. Vitebskiy, Waves Random Complex Media **16**, 293 (2006).
[24] J. P. Dowling, M. W. Davidson, M. J. Bloemer, and C. M. Bowden, J. Appl. Phys. **75**, 1896 (1994).
[25] J. Grgić, J. R. Ott, F. Wang, O. Sigmund, A.-P. Jauho, J. Mørk, and N. A. Mortensen, Phys. Rev. Lett. **108**, 183903 (2012).
[26] C. M. Bender, Rep. Prog. Phys. **70**, 947 (2007).
[27] S. Klaiman, U. Günther, and N. Moiseyev, Phys. Rev. Lett. **101**, 080402 (2008).
[28] C. E. Rüter, K. G. Makris, R. El-Ganainy, D. N. Christodoulides, M. Segev, and D. Kip, Nat. Phys. **6**, 192 (2010).
[29] M. Liertzer, L. Ge, A. Cerjan, A. D. Stone, H. E. Türeci, and S. Rotter, Phys. Rev. Lett. **108**, 173901 (2012).
[30] M. Brandstetter, M. Liertzer, C. Deutsch, P. Klang, J. Schöberl, H. E. Türeci, G. Strasser, K. Unterrainer, and S. Rotter, Nat. Commun. **5**, 4034 (2014).

[31] H. Hodaei, M.-A. Miri, M. Heinrich, D. N. Christodoulides, and M. Khajavikhan, Science **346**, 975 (2014).
[32] L. Feng, Z. J. Wong, R.-M. Ma, Y. Wang, and X. Zhang, Science **346**, 972 (2014).
[33] S. Wuestner, A. Pusch, K. L. Tsakmakidis, J. M. Hamm, and O. Hess, Phys. Rev. Lett. **105**, 127401 (2010).
[34] O. Hess, J. B. Pendry, S. A. Maier, R. F. Oulton, J. M. Hamm, and K. L. Tsakmakidis, Nat. Mater. **11**, 573 (2012).
[35] A. Marini and F. J. G. de Abajo, Sci. Rep. **6**, 20088 (2016).
[36] M. Dridi and G. C. Schatz, JOSA B **32**, 818 (2015).
[37] L. Xu, C. A. Curwen, P. W. C. Hon, Q.-S. Chen, T. Itoh, and B. S. Williams, Appl. Phys. Lett. **107**, 221105 (2015).
[38] L. Xu, D. Chen, T. Itoh, J. L. Reno, and B. S. Williams, Opt. Express **24**, 24117 (2016).
[39] M. A. K. Othman and F. Capolino, IEEE Microw. Wirel. Compon. Lett. **25**, 700 (2015).
[40] A. Figotin and I. Vitebskiy, Phys. Rev. B **67**, 165210 (2003).
[41] H. Ramezani, S. Kalish, I. Vitebskiy, and T. Kottos, Phys. Rev. Lett. **112**, 043904 (2014).
[42] A. Guo, G. J. Salamo, D. Duchesne, R. Morandotti, M. Volatier-Ravat, V. Aimez, G. A. Siviloglou, and D. N. Christodoulides, Phys. Rev. Lett. **103**, 093902 (2009).
[43] M. A. Othman, V. Galdi, and F. Capolino, Phys. Rev. B **95**, 104305 (2017).
[44] M. A. K. Othman and F. Capolino, IEEE Trans. Antennas Propag. **65**, 1 (2017).
[45] T. Kato, *Perturbation Theory for Linear Operators* (springer, 1995).
[46] A. Figotin and I. Vitebsky, Phys. Rev. E **63**, 066609 (2001).
[47] M. A. K. Othman, F. Yazdi, A. Figotin, and F. Capolino, Phys. Rev. B **93**, 024301 (2016).
[48] J. R. Burr, N. Gutman, C. M. de Sterke, I. Vitebskiy, and R. M. Reano, Opt. Express **21**, 8736 (2013).
[49] J. R. Burr and R. M. Reano, Opt. Express **23**, 30933 (2015).
[50] M. Y. Nada, M. A. K. Othman, O. Boyraz, and F. Capolino, J. Light. Technol. **(in press)**, (2018).
[51] S. Yarga, K. Sertel, and J. L. Volakis, IEEE Trans. Antennas Propag. **57**, 799 (2009).
[52] J. L. Volakis and K. Sertel, Proc. IEEE **99**, 1732 (2011).
[53] M. A. K. Othman, X. Pan, Y. Atmatzakis, and C. G. Christodoulou, IEEE Microw. Theory Tech. **1611**, 9 (2017).
[54] M. G. Wood, J. R. Burr, and R. M. Reano, Opt. Lett. **40**, 2493 (2015).
[55] M. G. Wood, J. R. Burr, and R. M. Reano, Opt. Express **24**, 23481 (2016).
[56] K. J. Vahala, Nature **424**, 839 (2003).
[57] M.-C. Tien, J. F. Bauters, M. J. R. Heck, D. T. Spencer, D. J. Blumenthal, and J. E. Bowers, Opt. Express **19**, 13551 (2011).
[58] G.-M. Parsanasab, M. Moshkani, and A. Gharavi, Opt. Express **23**, 8310 (2015).
[59] A. Yariv, Y. Xu, R. K. Lee, and A. Scherer, Opt. Lett. **24**, 711 (1999).
[60] A. Melloni, A. Canciamilla, C. Ferrari, F. Morichetti, L. O'Faolain, T. F. Krauss, R. D. L. Rue, A. Samarelli, and M. Sorel, IEEE Photonics J. **2**, 181 (2010).
[61] A. Canciamilla, M. Torregiani, C. Ferrari, F. Morichetti, R. M. D. L. Rue, A. Samarelli, M. Sorel, and A. Melloni, J. Opt. **12**, 104008 (2010).
[62] J.-P. Colinge, *Silicon-on-Insulator Technology - Materials to VLSI* (Springer, 1997).
[63] C. D. Meyer, *Matrix Analysis and Applied Linear Algebra* (Siam, 2000).
[64] A. Figotin and I. Vitebskiy, Waves Random Complex Media **16**, 293 (2006).
[65] M. A. Othman, M. Veysi, A. Figotin, and F. Capolino, Phys. Plasmas 1994-Present **23**, 033112 (2016).
[66] A. Figotin and I. Vitebskiy, Phys. Rev. E **72**, 036619 (2005).
[67] A. Figotin and I. Vitebskiy, Laser Photonics Rev. **5**, 201 (2011).
[68] Simon Ramo, John R. Whinnery, and Theodore Van Duzer, *Fields and Waves in Communication Electronics* (John Wiley & Sons, Inc., 1965).
[69] A. W. Snyder, JOSA **62**, 1267 (1972).
[70] A. Yariv, IEEE J. Quantum Electron. **9**, 919 (1973).
[71] A. Hardy and W. Streifer, J. Light. Technol. **3**, 1135 (1985).
[72] H. A. Haus and W. Huang, Proc. IEEE **79**, 1505 (1991).
[73] W.-P. Huang, JOSA A **11**, 963 (1994).
[74] M. K. Krage and G. I. Haddad, IEEE Trans. Microw. Theory Tech. **18**, 217 (1970).
[75] M. Othman, V. A. Tamma, and F. Capolino, IEEE Trans Plasma Sci **44**, 594 (2016).
[76] M. A. K. Othman, M. Veysi, A. Figotin, and F. Capolino, IEEE Trans. Plasma Sci. **44**, 918 (2016).
[77] K. Kurokawa, IEEE Trans. Microw. Theory Tech. **13**, 194 (1965).
[78] Clayton R. Paul, *Analysis of Multiconductor Transmission Lines*, 2nd Edition (Wiley-IEEE Press, 2007).
[79] K. Yee, IEEE Trans. Antennas Propag. **14**, 302 (1966).
[80] A. Taflove and S. C. Hagness, *Computational Electrodynamics: The Finite-Difference Time-Domain Method* (Artech House, 2000).
[81] Karl S. Kunz, *The Finite Difference Time Domain Method for Electromagnetics* (CRC Press, 1993).
[82] S.-H. Chang and A. Taflove, Opt. Express **12**, 3827 (2004).
[83] B. E. A. Saleh and M. C. Teich, in *Fundam. Photonics* (John Wiley & Sons, Inc., 1991), pp. 494–541.

[84] A. E. Siegman, *Lasers* (University Science Books, 1986).
[85] S. Campione, M. Albani, and F. Capolino, Opt. Mater. Express **1**, 1077 (2011).
[86] A. Yariv, Quantum Electron. IEEE J. Of **9**, 919 (1973).
[87] I. V. Barashenkov, L. Baker, and N. V. Alexeeva, Phys. Rev. A **87**, 033819 (2013).
[88] M. Y. Nada, M. A. Othman, and F. Capolino, Phys. Rev. B **96**, 184304 (2017).
[89] W. T. Silfvast, *Laser Fundamentals* (Cambridge University Press, 2004).
[90] D. Kajfez, *Multiconductor Transmission Lines* (Interaction Notes, Note 151, University of Mississippi. Department of Electrical Engineering, 1972).
[91] G. Miano and A. Maffucci, *Transmission Lines and Lumped Circuits: Fundamentals and Applications* (Academic Press, 2001).
[92] Leslie Hogben, *Handbook of Linear Algebra*, Second Edition (Chapman and Hall/CRC, 2013).
[93] W. K. K. L. B. Felsen, Proc. Symp. Millim. Waves J. Fox (1959).
[94] B. Wu, B. Sun, H. Xue, F. Xiao, Z. Huang, and X. Wu, Phys. B Condens. Matter **423**, 21 (2013).
[95] K. Yamada, in *Silicon Photonics II*, edited by D. J. Lockwood and L. Pavesi (Springer Berlin Heidelberg, 2011), pp. 1–29.
[96] Y. Dattner and O. Yadid-Pecht, IEEE Photonics J. **3**, 1123 (2011).
[97] K. F. Renk, *Basics of Laser Physics* (Springer Berlin Heidelberg, Berlin, Heidelberg, 2012).